\documentclass{article}
\usepackage{amssymb}
\usepackage{makeidx}
\usepackage{amsfonts}
\usepackage{amsmath}

\numberwithin{equation}{section}
\numberwithin{equation}{section} \textheight23cm \textwidth16cm
\newtheorem{theorem}{Theorem}

\newtheorem{example}[theorem]{Example}

\newtheorem{lemma}[theorem]{Lemma}

\input{tcilatex}
\begin{document}

\title{ Multi-component Painlev\'{e} ODE's and related non-autonomous KdV
stationary hierarchies}
\author{Maciej B\l aszak \\
%EndAName
Faculty of Physics, Department of Mathematical Physics and Computer
Modelling,\\
A. Mickiewicz University\\
Uniwersytetu Pozna\'nskiego 2, 61-614 Pozna\'{n}, Poland\\
\texttt{blaszakm@amu.edu.pl}}
\maketitle

\begin{abstract}
First, starting from two hierarchies of autonomous St\"{a}ckel ODE's, we
reconstruct the hierarchy of KdV stationary systems. Next, we deform
considered autonomous St\"{a}ckel systems to non-autonomous Painlev\'{e}
hierarchies of ODE's. Finally, we reconstruct the related non-autonomous KdV
stationary hierarchies from respective Painlev\'{e} systems.
\end{abstract}

\section{ Introduction}

Two particular classes of second order nonlinear ordinary differential
equations (ODE's) playing important role in a variety of branches of modern
mathematics and physics. The first class is represented by separable (St\"{a}%
ckel) equations with autonomous Hamiltonian representations. The second
class is represented by Painlev\`{e} equations with non-autonomous
Hamiltonian representations. Thus, both types of ODE's can be alternatively
considered as respective autonomous and non-autonomous Hamiltonian dynamical
systems. The St\"{a}ckel equations can be written in the so-called Lax
representation in the form of isospectral deformation equations while the
Painlev\`{e} equations can be written in the Lax representation in the form
of isomonodromic deformation equations. Both, separable and Painlev\`{e}
equations, appear in a wide range of applications in physics and
mathematics, so are definitely worth of investigation.

A significant progress in construction of new St\"{a}ckel and Painlev\'{e}
equations took place since the modern theory of nonlinear integrable PDE's
has been born (the so-called soliton theory). It was found that both type of
equations are inseparably connected with the soliton systems with whom they
share many properties. Actually, they have been constructed under particular
reductions of soliton PDE's.

The systematic construction of St\"{a}ckel systems of arbitrary degrees of
freedom from stationary flows and restricted flows of soliton hierarchies as
well as constrained flows of respective Lax hierarchies is nowadays well
developed \cite{bogov,ant1,ant2,Cao} (see also review of these methods in 
\cite{blaszak1998} and references therein). A bit less is known about
similar constructions of Painlev\'{e} systems with arbitrary number of
degrees of freedom. Nevertheless, many interesting results the reader can
find in \cite{Air,Chiba 2017,Clarkson 1999,Clarkson 2006,Fla,Gordoa,Koike
2008,Kudryahov 1997,Nou,Shimomura 2004,Takasaki}.

In this article we present an inverse approach to the relation between
Painlev\'{e} ODE's and soliton PDE's on the example of the KdV family.
Actually, from particular hierarchies of Painlev\'{e} systems we construct
the related hierarchies of non-autonomous and non-homogeneous deformations
of the KdV PDE's.

Let me briefly sketch this idea. Recently, we have developed a deformation
theory of autonomous St\"{a}ckel equations to non-autonomous Painlev\'{e}
equations \cite{Blaszak2021,BlaszakX1,BlaszakX2}. To be more precise, in the
literature was presented so far several constructions of Painlev\'{e}
hierarchies with increasing number of degrees of freedom. What important,
each number of degrees of freedom was related with a single equation. In
that sense, our deformation approach contains more. We start from a
hierarchy of autonomous separable systems with increasing number of degrees
of freedom, where a system of $n$ degrees of freedom consists $n$ commuting
(i.e. Frobenious integrable) evolution equations. After deformation we
obtain a hierarchy of non-autonomous Painlev\'{e} systems with increasing
number of degrees of freedom, where system of $n$ degrees of freedom
consists of $n$, Frobenious integrable, non-autonomous Painlev\'{e}
evolution equations. So, from now on, I will use the phrase a \emph{%
hierarchy of Painlev\'{e} systems} rather then a \emph{Painlev\'{e}
hierarchy }known from the literature.

On the other hand, in articles \cite{blaszak2008,blaszak2008a} we have made
an interesting observation that the complete soliton hierarchies can be
reconstructed from particular finite dimensional St\"{a}ckel systems,
representing their stationary flows.

In the following article we take the advantages from that observation and
taking the Painlev\'{e} deformations of St\"{a}ckel systems related to KdV
stationary flows, we construct related hierarchies of non-autonomous and
non-homogeneous deformations of the KdV hierarchy. Such new hierarchies
seems to be interesting objects for further investigation, as their
stationary flows reconstruct a particular hierarchies of Painlev\'{e}
systems.

This paper is organized as follows. In Section \ref{2} we briefly collect
the necessary information on the KdV hierarchy. In Section \ref{3} we
reconstruct the KdV hierarchy from the hierarchy of St\"{a}ckel systems,
which are representation of the KdV stationary systems related the first KdV
Hamiltonian structure. In Section \ref{4} we do the same from the hierarchy
of St\"{a}ckel systems which are representation of the KdV stationary
systems related the second KdV Hamiltonian structure. In Section \ref{5},
applying recently developed theory \cite{Blaszak2021,BlaszakX1,BlaszakX2},
we deform autonomous St\"{a}ckel hierarchies from Sections \ref{3} and \ref%
{4} to non-autonomous Painlev\'{e} hierarchies. Finally, in Section \ref{6},
we reconstruct, like in the autonomous case, two non-autonomous KdV
hierarchies of stationary systems, related to particular non-homogenous KdV
hierarchies.

\section{KdV hierarchy \label{2}}

Let us remind some elementary facts about the KdV hierarchy, important for
our further considerations. The KdV equation 
\begin{equation}
u_{t}=\tfrac{1}{4}u_{xxx}+\tfrac{3}{2}u\,u_{x}  \label{2.1}
\end{equation}%
is a member of the following bi-Hamiltonian chain of nonlinear PDE's%
\begin{equation}
u_{t_{n}}=\mathcal{K}_{n}=\pi _{0}d\mathcal{H}_{n}=\pi _{1}d\mathcal{H}%
_{n-1},\ \ \ \ n=1,2,...  \label{2.2}
\end{equation}%
where two Poisson operators are 
\begin{equation}
\pi _{0}=\partial _{x},\ \ \ \pi _{1}=\tfrac{1}{4}\partial _{x}^{3}+\tfrac{1%
}{2}u\partial _{x}+\tfrac{1}{2}\partial _{x}u.  \label{2.3}
\end{equation}%
The hierarchy (\ref{2.2}) can be generated by a recursion operator and its
adjoint in a following way%
\begin{equation}
N=\pi _{1}\pi _{0}^{-1}=\tfrac{1}{4}\partial _{x}^{2}+u+\tfrac{1}{2}%
u_{x}\partial _{x}^{-1},\ \ \ N^{\dagger }=\tfrac{1}{4}\partial _{x}^{2}+u-%
\tfrac{1}{2}\partial _{x}^{-1}u_{x},  \label{2.4}
\end{equation}%
\begin{equation}
\mathcal{K}_{n+1}=N^{n}\mathcal{K}_{1},\ \ \ \ \ \ \gamma _{n}=d\mathcal{H}%
_{n}=\left( N^{\dagger }\right) ^{n}\gamma _{0},\ \ \ \ n=1,2,.....
\label{2.5}
\end{equation}%
In particular, conserved one-forms are 
\begin{align}
\gamma _{0}& =2, & &  \notag \\
\gamma _{1}& =u, & &  \notag \\
\gamma _{2}& =\tfrac{1}{4}u_{xx}+\tfrac{3}{4}u^{2}, & &  \label{2.5a} \\
\gamma _{3}& =\tfrac{1}{16}u_{4x}+\tfrac{5}{8}u\,u_{xx}+\tfrac{5}{16}%
u_{x}^{2}+\frac{5}{8}u^{3}, & &  \notag \\
\gamma _{4}& =\tfrac{1}{64}u_{6x}+\tfrac{7}{32}u\,u_{4x}+\tfrac{7}{16}%
u_{x}u_{3x}+\tfrac{21}{64}u_{xx}^{2}+\tfrac{35}{32}u^{2}u_{xx}+\tfrac{35}{32}%
u\,u_{x}^{2}+\tfrac{35}{64}u^{4}, & &  \notag \\
& \vdots & &  \notag
\end{align}%
and related symmetries are 
\begin{align}
\mathcal{K}_{1}& =u_{x}, & &  \notag \\
\mathcal{K}_{2}& =\tfrac{1}{4}u_{xxx}+\tfrac{3}{2}u\,u_{x}, & &  \label{2.5b}
\\
\mathcal{K}_{3}& =\tfrac{1}{16}u_{5x}+\tfrac{5}{8}u\,u_{3x}+\tfrac{5}{4}%
u_{x}u_{xx}+\frac{15}{8}u^{2}u_{x}, & &  \notag \\
\mathcal{K}_{4}& =\tfrac{1}{64}u_{7x}+\tfrac{7}{32}u\,u_{5x}+\tfrac{21}{32}%
u_{x}u_{4x}+\tfrac{35}{32}u_{xx}u_{3x}+\tfrac{35}{32}u_{x}^{3}+\tfrac{35}{8}%
u\,u_{x}u_{xx}+\tfrac{35}{32}u^{2}u_{3x}+\tfrac{35}{16}u^{3}u_{x}, & & 
\notag \\
& \vdots & &  \notag
\end{align}%
As $u$ belongs to the whole hierarchy (\ref{2.2}) it depends on infinitely
many evolution parameters: $u=u(t_{1},t_{2},t_{3},...)$.

In addition, with the KdV hierarchy of symmetries is related a hierarchy of
master symmetries 
\begin{equation}
\sigma _{m}=N^{m+1}\sigma _{-1},\ \ \ \ \ \tau _{-1}=1,  \label{2.6a}
\end{equation}%
non-local in general 
\begin{align}
\sigma _{-1}& =1, & &  \notag \\
\sigma _{0}& =u+\tfrac{1}{2}xu_{x}, & &  \label{2.6aa} \\
\sigma _{1}& =\tfrac{1}{2}u_{xx}+\tfrac{1}{8}xu_{3x}+u^{2}+\tfrac{1}{2}%
xu\,u_{x}+\tfrac{1}{4}u_{x}\partial _{x}^{-1}u, & &  \notag \\
& \vdots & &  \notag
\end{align}%
Both, symmetries $\mathcal{K}_{n}$ (\ref{2.5}) and master symmetries $\sigma
_{m}$ (\ref{2.6a}) constitute so called Virasoro algebra (hereditary
algebra) 
\begin{equation}
\lbrack \mathcal{K}_{m},\mathcal{K}_{n}]=0,\ \ \ [\mathcal{\sigma }_{m},%
\mathcal{K}_{n}]=(n-\tfrac{1}{2})\mathcal{K}_{n+m},\ \ \ \ [\sigma
_{m},\sigma _{n}]=(n-m)\sigma _{n+m}.  \label{2.6b}
\end{equation}

Alternatively, the hierarchy (\ref{2.2}) can be reconstructed from the
isospectral Lax representation. Actually, consider some eigenvalue problem
together with time evolutions of its eigenfunctions%
\begin{equation}
\begin{array}{l}
L(u)\psi =\lambda \psi ,\ \ \ \ \ \ \ \ \lambda _{t_{n}}=0, \\ 
\psi _{t_{n}}=B_{n}(u)\psi ,\ \ \ \ n=1,2,...,%
\end{array}
\label{L1}
\end{equation}%
where $L$ and $B_{r}$ are some differential operators. The compatibility
conditions (Frobenius integrability conditions) for (\ref{L1}) takes the
form 
\begin{equation}
L_{t_{n}}=[B_{n},L],\ \ \ \ \ \ \ n=1,2,...,  \label{L2}
\end{equation}%
known as isospectral deformation equations, as the eigenvalues of the
operator $L$ are independent of all times $t_{r}$, and are equivalent with
evolutionary hierarchy of PDE's. For the KdV hierarchy 
\begin{equation}
L=\partial _{x}^{2}+u,\ \ \ \ \ \ \ \ B_{n}=\left( L^{n-\frac{1}{2}}\right)
_{\geq 0}=\sum_{i=0}^{n-1}\left( -\frac{1}{4}\mathcal{K}_{i}+\frac{1}{2}%
\gamma _{i}\partial _{x}\right) L^{n-i-1}\ \   \label{2.6}
\end{equation}%
where explicitly 
\begin{align}
B_{1}& =\partial _{x}, & &  \notag \\
B_{2}& =\partial _{x}^{3}+\frac{3}{2}u\partial _{x}+\frac{3}{4}u_{x}, & &
\label{L3} \\
B_{3}& =\partial _{x}^{5}+\tfrac{5}{2}u\partial _{x}^{3}+\tfrac{15}{4}%
u_{x}^{2}\partial _{x}^{2}+\tfrac{5}{8}(3u^{2}+5u_{xx})\partial _{x}+\tfrac{%
15}{16}(u_{3x}+2u\,u_{x}), & &  \notag \\
& \vdots & &  \notag
\end{align}

Isospectral deformation equations (\ref{L2}), (\ref{2.6}) can be presented
in the equivalent form of so called zero curvature equations, more suitable
for our further considerations. Rewriting equations (\ref{L1}) for the KdV
hierarchy in the form%
\begin{align}
\Psi _{x} &=U(\lambda ;u)\Psi ,\ \ \ \ \ \ \Psi =(\psi ,\psi _{x})^{T}
\label{L4a} \\
\Psi _{t_{n}} &=V_{n}(\lambda ;u)\Psi ,\ \ \ \ n=1,2,...,  \label{L4b}
\end{align}%
the compatibility conditions for (\ref{L4a}) and (\ref{L4b}) takes the form 
\begin{equation}
\frac{d}{dt_{n}}U-\frac{d}{dx}V_{n}+[U,V_{n}]=0\Longleftrightarrow u_{t_{n}}=%
\mathcal{K}_{n}\ \ \ \ n=1,2,...\text{,}  \label{L5}
\end{equation}%
known as zero curvature conditions and reconstruct the KdV hierarchy. In (%
\ref{L5}) $\frac{d}{dx}$ means the total $x$-derivative and $\frac{d}{dt_{n}}
$ means the $t_{n}$-evolutionary derivative. Actually, we have%
\begin{equation}
V_{k}=\left( 
\begin{array}{cc}
-\frac{1}{2}P_{x} & P \\ 
P(\lambda -u)-\frac{1}{2}P_{xx} & \frac{1}{2}P_{x}%
\end{array}%
\right) ,\ \ \ P_{k}=\frac{1}{2}\sum_{i=0}^{k-1}\gamma _{i}\lambda ^{k-i-1}.
\label{L6}
\end{equation}%
\ In consequence%
\begin{equation}
V_{1}=U=\left( 
\begin{array}{cc}
0 & 1 \\ 
\lambda -u & 0%
\end{array}%
\right) ,\ \ \ V_{2}=\left( 
\begin{array}{cc}
-\frac{1}{4}u_{x} & \lambda +\frac{1}{2}u \\ 
\lambda ^{2}-\frac{1}{2}u\lambda -\frac{1}{2}u^{2}-\frac{1}{4}u_{xx} & \frac{%
1}{4}u_{x}%
\end{array}%
\right) ,  \label{L7}
\end{equation}%
\begin{equation*}
\end{equation*}%
\begin{equation}
V_{3}=\left( 
\begin{array}{cc}
-\frac{1}{4}u_{x}\lambda -\frac{1}{16}(u_{3x}+6u\,u_{x}) & \lambda ^{2}+%
\frac{1}{2}u\lambda +\frac{1}{8}(u_{xx}+3u^{2}) \\ 
\lambda ^{3}-\frac{1}{2}u\lambda ^{2}-\frac{1}{8}(u_{xx}+u^{2})\lambda -(%
\frac{1}{16}u_{4x}+\frac{1}{2}u\,u_{xx}+\frac{3}{8}u_{x}^{2}+\frac{3}{8}%
u^{3}) & \frac{1}{4}u_{x}\lambda +\frac{1}{16}(u_{3x}+6u\,u_{x})%
\end{array}%
\right) ,  \label{L8}
\end{equation}%
\begin{equation*}
\vdots
\end{equation*}

Finally, the compatibility conditions between equations from (\ref{L4b})
take the form of zero curvature conditions 
\begin{equation}
\frac{d}{dt_{r}}V_{s}-\frac{d}{dt_{s}}V_{r}+[V_{s},V_{r}]=0,\ \ \ \ \
r,s=1,2,...  \label{L9}
\end{equation}%
and are valid as differential consequences of commutativity of $r$-th and $s$%
-th KdV vector fields.

Let us define the $n$-th KdV stationary systems as \cite{BS} 
\begin{equation}
u_{t_{r}}=\mathcal{K}_{r},\ \ \ 0=\mathcal{K}_{n+1},\ \ \ r=1,...,n.
\label{s1}
\end{equation}%
The stationary restriction $0=\mathcal{K}_{n+1}$ provides constraint on the
infinite-dimensional (functional) manifold, on which the KdV hierarchy is
defined, reducing it to the finite-dimensional (stationary) submanifold $%
\mathcal{M}_{n}$ of dimension $(2n+1)$, parametrized by jet coordinates $%
u,u_{x},\ldots ,u_{(2n)x}$. The system (\ref{s1}) has two different $2n$
dimensional representations, related to two different foliations of $%
\mathcal{M}_{n}$. The first one is of the form 
\begin{equation}
u_{t_{r}}=\mathcal{K}_{r},\ \ \ 0=\gamma _{n+1}+c,\ \ \ \ \ r=1,...,n,\ \ \
\ \ \ c\in \mathbb{R},  \label{s2}
\end{equation}%
where last equation in (\ref{s2}) is integrated an $(n+1)$ stationary flow
of the KdV hierarchy (\ref{2.2}) in the first Hamiltonian representation 
\begin{equation}
0=\mathcal{K}_{n+1}=\partial _{x}\gamma _{n+1}\Longrightarrow \ 0=\gamma
_{n+1}+c.  \label{s3}
\end{equation}%
The second one is 
\begin{equation}
\ u_{t_{r}}=\mathcal{K}_{r},\ \ \ \ 0=\tfrac{1}{2}\gamma _{n}(\gamma
_{n})_{xx}-\tfrac{1}{4}[(\gamma _{n})_{x}]^{2}+u\gamma _{n}^{2}+c,\ \ \ \
r=1,...,n,  \label{s4}
\end{equation}%
where the last equation in (\ref{s4}) is integrated an $(n+1)$ stationary
flow of the hierarchy (\ref{2.2}) in the second Hamiltonian representation.
Indeed, differentiating it by $x$ and dividing by $2\gamma _{n}$ we get 
\begin{equation}
0=\left( \tfrac{1}{4}\partial _{x}^{3}+\tfrac{1}{2}u\partial _{x}+\tfrac{1}{2%
}\partial _{x}u\right) \gamma _{n}=\mathcal{K}_{n+1}.  \label{s5}
\end{equation}%
Both representations constitute systems of ODE's with $n$ degrees of
freedom, on $2n$-dimensional phase space with jet coordinates $%
u,u_{x},\ldots ,u_{(2n-1)x},$ as higher derivatives of $u$ with respect to $%
x $ are eliminated by the constraint (\ref{s3}) and (\ref{s4}), respectively.

Moreover, Lax representation of the system (\ref{s2}) is given by 
\begin{equation}
\frac{d}{dt_{r}}V_{n+1}=[V_{r},V_{n+1}],\ \ \ r=2,...,n,\ \ \ \ \ \ \frac{d}{%
dx}V_{n+1}=[V_{1},V_{n+1}],  \label{s7}
\end{equation}%
following from (\ref{L9}), with constraint $0=\gamma _{n+1}+c$, imposed on $%
V_{n+1}$, while Lax representation of the system (\ref{s4}) is given by (\ref%
{s7}) with constraint $0=\tfrac{1}{2}\gamma _{n}(\gamma _{n})_{xx}-\tfrac{1}{%
4}(\gamma _{n})_{x}^{2}+u\gamma _{n}^{2}+c$, imposed on $V_{n+1}$.

St\"{a}ckel representations of stationary systems (\ref{s2}) and (\ref{s4})
were derived in \cite{BS} and are presented in the next two sections.

\section{Hamiltonian representation of the first KdV hierarchy of stationary
systems \label{3}}

Let us briefly systematize known facts about the hierarchy of St\"{a}ckel
systems, being the Hamiltonian representation of the KdV hierarchy of
stationary systems (\ref{s2}) \cite{ant1,blaszak1998,blaszak2008,BS}.

Consider finite dimensional separable Hamiltonian systems, so called St\"{a}%
ckel systems, generated by the following hyperelliptic separation (spectral)
curves on $(\lambda ,\mu )$-plane 
\begin{equation}
\lambda ^{2n+1}+c\lambda ^{n}+\sum_{r=1}^{n}h_{r}\lambda ^{n-r}=\mu ^{2},\ \
\ \ \ n\in \mathbb{N},\ \ \ \ c=const.  \label{2.7}
\end{equation}%
For fix $n\in \mathbb{N}$, consider a $2n$-dimensional Poisson manifold $%
(M,\pi )$ and a particular set $\xi =(\lambda _{1},\ldots ,\lambda
_{n},\allowbreak \mu _{1},\ldots ,\mu _{n})\in M$ of Darboux (canonical)
coordinates on $M$, i.e. $\{\lambda _{i},\lambda _{j}\}_{\pi }=\{\mu
_{i},\mu _{j}\}_{\pi }=0$, $\{\lambda _{i},\mu _{j}\}_{\pi }=\delta _{ij}$.
By taking $n$ copies of (\ref{2.7}) at points $(\lambda ,\mu )=(\lambda
_{i},\mu _{i})$, $i=1,\dotsc ,n$, we obtain a system of $n$ linear equations
(separation relations) for $h_{r}$%
\begin{equation}
\lambda _{i}^{2n+1}+c\lambda _{i}^{n}+\sum_{r=1}^{n}h_{r}\lambda
_{i}^{n-k}=\mu _{i}^{2},\ \ \ \ \ i=1,...,n,\ \ \ \ n\in \mathbb{N}.
\label{2.8}
\end{equation}%
Solving (\ref{2.8}) with respect to $h_{r}$ yields $n$ functions
(Hamiltonians) $h_{r}$ on $(M,\pi )$ 
\begin{eqnarray}
h_{r} &=&\sum_{i=1}^{n}(-1)^{r+1}\frac{\partial s_{r}}{\partial \lambda _{i}}%
\frac{\mu _{i}^{2}}{\Delta _{i}}+\sum_{i=1}^{n}(-1)^{r}\frac{\partial s_{r}}{%
\partial \lambda _{i}}\frac{\lambda _{i}^{2n+1}+c\lambda _{i}^{n}}{\Delta
_{i}}  \notag \\
&&  \label{2.9} \\
&=&\mu ^{T}K_{r}G_{0}\mu +V_{r}^{(2n+1)}+cV_{r}^{(n)},\quad r=1,\ldots ,n 
\notag
\end{eqnarray}%
where%
\begin{equation}
G_{0}=\text{diag}\left( \frac{1}{\Delta _{1}},\ldots ,\frac{1}{\Delta _{n}}%
\right) ,\ \ \ \ \ \ \Delta _{j}=\prod\limits_{k\neq j}(\lambda _{j}-\lambda
_{k}),  \label{2.9a}
\end{equation}%
\begin{equation}
K_{r}=(-1)^{r+1}\text{diag}\left( \frac{\partial s_{r}}{\partial \lambda _{1}%
},\cdots ,\frac{\partial s_{r}}{\partial \lambda _{n}}\right) ,\quad
r=1,\ldots ,n,  \label{2.9b}
\end{equation}%
$s_{r}$ are elementary symmetric polynomials in $\lambda _{i}$ and $%
V_{r}^{(k)},\ k\in \mathbb{Z}$ are basic separable potentials. Matrix $G_{0}$
can be interpreted as a contravariant metric tensor on an $n$-dimensional
manifold $Q$ such that $M=T^{\ast }Q$. It can be shown that the metric $%
G_{0} $ is flat. Matrices $K_{r}$ are $(1,1)$-Killing tensors for the metric 
$G_{0} $.

Thus, we have constructed from scratch infinite hierarchy of separable
autonomous Hamiltonian systems 
\begin{equation}
\xi _{t_{r}}=\pi dh_{r}=X_{r},\ \ \ \ r=1,...,n,\ \ \ \ \ n\in \mathbb{N},
\label{2.9c}
\end{equation}%
where each system consists of $n$ evolution ODE's. The first Hamiltonian $%
h_{1}$ of each system can be interpreted as the Hamiltonian of a particle in
the pseudo-Riemannian $n$-dimensional configuration space $%
(Q,g_{0}=G_{0}^{-1}),$ moving under particular potential force and the
remaining Hamiltonians are its constants of motion.

By their very construction from separation relations, the Hamiltonians $%
h_{r} $ Poisson commute for all $r,k=1,\ldots ,n$ 
\begin{equation}
\{h_{r},h_{k}\}_{\pi }=\pi (dh_{r},dh_{k})=0,  \label{2.10}
\end{equation}%
and so that $[X_{r},X_{k}]=0.$ It means that each system is Liouville
integrable and hence is Frobenius integrable, i.e. the system (\ref{2.9c})
has a common, unique solution $\xi (t_{1},\ldots ,t_{n},\xi _{0})$ through
each point $\xi _{0}\in M,$ depending in general on all the evolution
parameters $t_{s},$ constructed from separation relations (\ref{2.8}) by
quadratures.

In what follows we will work in so called Vi\'{e}te canonical coordinates $%
(q,p)\in T^{\ast }Q$ 
\begin{equation}
q_{i}=(-1)^{i}s_{i}(\lambda ),\quad p_{i}=\ -\sum_{k=1}^{n}\frac{\lambda
_{k}^{n-i}\mu _{k}}{\Delta _{k}},\quad i=1,\dotsc ,n  \label{2.11}
\end{equation}%
in which all functions $h_{r}$ are polynomial functions of their arguments.
Explicitly 
\begin{equation}
G_{0}^{ij}=q_{i+j-n-1},\ \ \ \ \left( K_{r}\right) _{j}^{i}=%
\begin{cases}
q_{i-j+r-1}, & \text{$i\leq j$ and $r\leq j$} \\ 
-q_{i-j+r-1}, & \text{$i>j$ and $r>j$} \\ 
0, & \text{otherwise}%
\end{cases}
\label{2.11aa}
\end{equation}%
where we set $q_{0}=1$, $q_{k}=0$ for $k<0$ or $k>n$. Elementary separable
potentials $V_{r}^{(\alpha )}$ can be explicitly constructed by the
recursion formula \cite{blaszak2011} 
\begin{equation}
V^{(\alpha )}=R^{\alpha }V^{(0)},\qquad V^{(\alpha )}=(V_{1}^{(\alpha
)},\dotsc ,V_{n}^{(\alpha )})^{T},\ \ \ \ \ R=%
\begin{pmatrix}
-q_{1} & \ 1 & \ 0 & \ 0 \\ 
\vdots & \ 0 & \ \ddots & \ 0 \\ 
\vdots & \ 0 & \ 0 & \ 1 \\ 
-q_{n} & \ 0 & \ 0 & \ 0%
\end{pmatrix}%
,  \label{2.11c}
\end{equation}%
with $V^{(0)}=(0,\dotsc ,0,-1)^{T}$. The first $n$ basic separable
potentials are trivial%
\begin{equation*}
V_{k}^{(\alpha )}=-\delta _{k,n-\alpha },\quad \alpha =0,\dotsc ,n-1.
\end{equation*}%
The first nontrivial positive and negative potentials are 
\begin{equation*}
V^{(n)}=(q_{1},\dotsc ,q_{n})^{T},\ \ \ \ V^{(-1)}=\left( \frac{1}{q_{n}}%
,\dotsc ,\frac{q_{n-1}}{q_{n}}\right) ^{T}
\end{equation*}%
and higher positive and negative potentials are more complicated polynomials
and rational functions in all $q_{i}$.

Besides, the autonomous Hamiltonian equations (\ref{2.9c}) are represented
by (i.e. are differential consequences of) Lax isospectral deformation
equations 
\begin{equation}
\frac{d}{dt_{r}}L(\lambda ;\xi )=[U_{r}(\lambda ;\xi ),L(\lambda ;\xi
)],\quad r=1,\ldots ,n,  \label{4.14}
\end{equation}%
where $\frac{d}{dt_{r}}$\ is the evolutionary derivative along the $r$-th
flow from (\ref{2.9c})%
\begin{equation}
\frac{d}{dt_{r}}A=\frac{\partial A}{dt_{r}}+\{A,h_{r}\}=\frac{\partial A}{%
\partial t_{r}}+A^{\prime }[X_{r}],  \label{4.14a}
\end{equation}
with $L(\lambda ;\xi )$ and $U_{r}(\lambda ;\xi )$ being matrices belonging
to some Lie algebra and depending rationally on the spectral parameter $%
\lambda $.

Actually, for the hierarchy generated by Hamiltonians (\ref{2.9}), there is
an infinitely many nonequivalent Lax representations \cite{Blaszak2019a}. \
Here we chose the one compatible with (\ref{L5})-(\ref{L8}).

\begin{theorem}
\cite{Blaszak2019a} The $L(\lambda ;\xi )$ matrix for the system (\ref{2.9c}%
) takes a form \ 
\begin{equation}
L(\lambda ;\xi )=\left( 
\begin{array}{cc}
v(\lambda ;\xi ) & u(\lambda ;\xi ) \\ 
w(\lambda ;\xi ) & -v(\lambda ;\xi )%
\end{array}%
\right) ,  \label{2.20}
\end{equation}%
where, in Vi\'{e}te coordinates $\xi =(q,p)$, we have%
\begin{equation}
u(\lambda ;q)=\lambda ^{n}+\sum_{k=1}^{n}q_{k}\lambda ^{n-k},\ \ \ v(\lambda
;q,p)=-\sum_{k=1}^{n}M_{k}(q,p)\lambda ^{n-k},\ \
M_{k}=\sum_{j=1}^{n}G_{0}^{kj}p_{j}  \label{2.21}
\end{equation}%
and 
\begin{equation}
w(\lambda ;q,p,t)=-\left[ \frac{v(\lambda )^{2}-\lambda ^{2n+1}-c\lambda ^{n}%
}{u(\lambda )}\right] _{+}.  \label{2.22}
\end{equation}%
Here, the operation $[\cdot ]_{+}$ means the projection on the uniquely
defined quotient of the division of an analytic function $A$ over a (pure)
polynomial $u(\lambda )$ such that the following decomposition holds: 
\begin{equation*}
A=\left[ \frac{A}{u}\right] _{+}u+r,
\end{equation*}%
where the (unique) remainder $r$ is a lower degree polynomial than the
polynomial $u$, see for details \cite{Blaszak2019a}. In particular, when $A$
is a Laurent polynomial, we have 
\begin{equation*}
\left[ \frac{A}{u}\right] _{+}\equiv \left[ \frac{\lbrack A]_{\geqslant 0}}{u%
}\right] _{\geqslant 0}+\left[ \frac{[A]_{<0}}{u}\right] _{<0},
\end{equation*}%
where $[\cdot ]_{\geqslant 0}$ is the projection on the part consisting of
non-negative degree terms in the expansion into Laurent series at $\infty $
and $[\cdot ]_{<0}$ is the projection on the part consisting of negative
degree terms in the expansion into Laurent series at $0$. Moreover, 
\begin{equation}
U_{r}(\lambda ;\xi )=\left[ \frac{u_{r}(\lambda )L(\lambda )}{u(\lambda )}%
\right] _{+},\quad \text{for $r=1,\dotsc ,n$},  \label{2.23}
\end{equation}%
where 
\begin{equation*}
u_{r}(\lambda )=\lambda ^{r-1}+\sum_{k=1}^{r-2}q_{k}\lambda ^{r-k-1}.
\end{equation*}
\end{theorem}

What is interesting, the separation curve (\ref{2.7}) is reconstructed from
Lax matrix $L(\lambda ;\xi )$ through 
\begin{equation*}
\det [L(\lambda ;\xi )-\mu I]=0.
\end{equation*}

The relation between the St\"{a}ckel hierarchy (\ref{2.9c}) and the KdV
stationary hierarchy (\ref{s2}) is as follows.

\begin{theorem}
\cite{BS} For fixed $n\in \mathbb{N}$ and identification of $t_{1}$ with $x$
we get the following equivalence between the St\"{a}ckel system (\ref{2.9c})
and the $n$-th KdV stationary system (\ref{s2}) 
\begin{equation}
\xi _{t_{r}}=X_{r},\ \ \ \ \ r=1,...,n  \label{2.13a}
\end{equation}%
\begin{equation*}
\Updownarrow
\end{equation*}%
\begin{equation}
u_{t_{r}}=\mathcal{K}_{r},\ \ \ \ 0=\gamma _{n+1}+c,\ \ \ \ r=1,...,n,
\label{2.13b}
\end{equation}%
where the transformation between jet and Vi\'{e}te coordinates is as follows 
\begin{equation}
q_{k}=\tfrac{1}{2}\gamma _{k},\ \ \ \ k=1,...,n,\ \ \ \ \ \ p_{k}=\frac{1}{2}%
\sum_{j=1}^{n}(G_{0}^{-1})_{kj}(q_{j})_{x}.  \label{2.15}
\end{equation}%
In particular, the first flow $\xi _{x}=X_{1}$ reconstructs (\ref{2.15}) and
the constraint 
\begin{equation}
0=(p_{1})_{x}-(X_{1})^{n+1}\Leftrightarrow 0=\gamma _{n+1}+c.  \label{2.15c}
\end{equation}%
Besides, the first component of the $r$-th flow (\ref{2.13a}) reconstructs
the $r$-th KdV equation 
\begin{equation}
(q_{1})_{t_{r}}=(X_{r})^{1}\Leftrightarrow u_{t_{r}}=\mathcal{K}_{r},\ \ \ \
r=1,...,n  \label{2.15b}
\end{equation}%
while the remaining components of systems (\ref{2.13a}) for $r=2,...,n$ are
differential consequences of (\ref{2.15c}) and (\ref{2.15b}). On the level
of Lax representation of the St\"{a}ckel hierarchy (\ref{4.14}) and the Lax
representation (\ref{s7}) of the KdV stationary system (\ref{s2}), fixing $n$
we have the following identities 
\begin{equation}
U_{i}=V_{i},\ \ \ \ i=1,...,n,  \label{2.15a}
\end{equation}%
and additionally 
\begin{equation}
L=V_{n+1}\text{ \ \ \ under constraint \ }0=\gamma _{n+1}+c.  \label{2.15aa}
\end{equation}
\end{theorem}

Both systems (\ref{2.13a}) and (\ref{2.13b}) share the same solutions, which
are the so called finite gape solutions and rational solutions of related
KdV hierarchy \cite{blaszak2008}.

\begin{example}
\label{e1}Consider the case $n=3.$ In $(q,p)$ coordinates, three
Hamiltonians $h_{r}=p^{T}K_{r}G_{0}p+V_{r}^{(7)}+cV_{r}^{(3)}$, metric
tensor $G_{0}$ and its inverse are%
\begin{eqnarray*}
h_{1}
&=&(2p_{1}p_{3}+p_{2}^{2}+2q_{1}p_{2}p_{3}+q_{2}p_{3}^{2})+(q_{1}^{5}-4q_{1}^{3}q_{2}+3q_{1}^{2}q_{3}+3q_{1}q_{2}^{2}-2q_{2}q_{3})+cq_{1},
\\
h_{2}
&=&[2q_{1}p_{1}p_{3}+2q_{1}p_{2}^{2}+2p_{1}p_{2}+(q_{1}q_{2}-q_{3})p_{3}^{2}+2q_{1}^{2}p_{2}p_{3}]+(q_{1}^{4}q_{2}-q_{1}^{3}q_{3}-3q_{1}^{2}q_{2}^{2}+4q_{1}q_{2}q_{3}+q_{2}^{3}-q_{3}^{2})+cq_{2},
\\
h_{3}
&=&p_{1}^{2}+2q_{1}p_{1}p_{2}+2q_{2}p_{1}p_{3}+q_{1}^{2}p_{2}^{2}+(q_{2}^{2}-q_{1}q_{3})p_{3}^{2}+2(q_{1}q_{2}-q_{3})p_{2}p_{3}
\\
&&+(q_{1}^{4}q_{3}-3q_{1}^{2}q_{2}q_{3}+2q_{1}q_{3}^{2}+q_{2}^{2}q_{3})+cq_{3},
\end{eqnarray*}
\begin{equation*}
G_{0}=\left( 
\begin{array}{ccc}
0 & 0 & 1 \\ 
0 & 1 & q_{1} \\ 
1 & q_{1} & q_{2}%
\end{array}%
\right) ,\ \ \ \ G_{0}^{-1}=\left( 
\begin{array}{ccc}
q_{1}^{2}-q_{2} & -q_{1} & 1 \\ 
-q_{1} & 1 & 0 \\ 
1 & 0 & 0%
\end{array}%
\right) .
\end{equation*}%
The related autonomous evolution equations are%
\begin{equation}
\left( 
\begin{array}{c}
q_{1} \\ 
q_{2} \\ 
q_{3} \\ 
p_{1} \\ 
p_{2} \\ 
p_{3}%
\end{array}%
\right) _{x}=X_{1}=\left( 
\begin{array}{l}
\ \ 2p_{3} \\ 
\ \ 2q_{1}p_{3}+2p_{2} \\ 
\ \ 2q_{1}p_{2}+2q_{2}p_{3}+2p_{1} \\ 
-2p_{2}p_{3}-5q_{1}^{4}+12q_{1}^{2}q_{2}-6q_{1}q_{3}-3q_{2}^{2}-c \\ 
-p_{3}^{2}+4q_{1}^{3}-6q_{1}q_{2}+2q_{3} \\ 
-3q_{1}^{2}+2q_{2}%
\end{array}%
\right)   \label{2.16}
\end{equation}%
\begin{equation}
\left( 
\begin{array}{c}
q_{1} \\ 
q_{2} \\ 
q_{3} \\ 
p_{1} \\ 
p_{2} \\ 
p_{3}%
\end{array}%
\right) _{t_{2}}=X_{2}=\left( 
\begin{array}{l}
\ \ 2p_{2}+2q_{1}p_{3} \\ 
\ \ 2p_{1}+2q_{1}^{2}p_{3}+4q_{1}p_{2} \\ 
\ \ 2q_{1}p_{1}+2(q_{1}q_{2}-q_{3})p_{3}+2q_{1}^{2}p_{2} \\ 
-2p_{1}p_{3}-2p_{2}^{2}-4q_{1}p_{2}p_{3}-q_{2}p_{3}^{2}+6q_{1}q_{2}^{2}-4q_{1}^{3}q_{2}+3q_{1}^{2}q_{3}-4q_{2}q_{3}
\\ 
-q_{1}p_{3}^{2}-q_{1}^{4}+6q_{1}^{2}q_{2}-4q_{1}q_{3}-3q_{2}^{2}-c \\ 
\ \ p_{3}^{2}+q_{1}^{3}-4q_{1}q_{2}+2q_{3}%
\end{array}%
\right)   \label{2.17}
\end{equation}%
\begin{equation}
\left( 
\begin{array}{c}
q_{1} \\ 
q_{2} \\ 
q_{3} \\ 
p_{1} \\ 
p_{2} \\ 
p_{3}%
\end{array}%
\right) _{t_{3}}=X_{3}=\left( 
\begin{array}{l}
\ \ 2q_{1}p_{2}+2q_{2}p_{3}+2p_{1} \\ 
\ \ 2q_{1}^{2}p_{2}+2q_{1}p_{1}+2(q_{1}q_{2}-q_{3})p_{3} \\ 
\ \ 2q_{2}p_{1}+2(q_{2}^{2}-q_{1}q_{3})p_{3}+2(q_{1}q_{2}-q_{3})p_{2} \\ 
-2q_{1}p_{2}^{2}-2q_{2}p_{2}p_{3}+q_{3}p_{3}^{2}-4q_{1}^{3}q_{3}+6q_{1}q_{2}q_{3}-2p_{1}p_{2}-2q_{3}^{2}
\\ 
-2p_{1}p_{3}-2q_{1}p_{2}p_{3}-2q_{2}p_{3}^{2}+3q_{1}^{2}q_{3}-2q_{2}q_{3} \\ 
\ \
q_{1}p_{3}^{2}+3q_{1}^{2}q_{2}+2p_{2}p_{3}-q_{1}^{4}-4q_{1}q_{3}-q_{2}^{2}-c%
\end{array}%
\right) .  \label{2.18}
\end{equation}%
Lax representations (\ref{4.14}) of considered equations are as follows 
\begin{equation}
L=\left( 
\begin{array}{cc}
\begin{array}{c}
-p_{3}\lambda ^{2}-(q_{1}p_{3}+p_{2})\lambda -p_{1}-q_{1}p_{2}-q_{2}p_{3} \\ 
\left. {}\right. 
\end{array}
& 
\begin{array}{c}
\lambda ^{3}+q_{1}\lambda ^{2}+q_{2}\lambda +q_{3} \\ 
\left. {}\right. 
\end{array}
\\ 
\begin{array}{c}
\lambda ^{4}-q_{1}\lambda ^{3}-(q_{2}-q_{1}^{2})\lambda
^{2}-(p_{3}^{2}+q_{1}^{3}-2q_{1}q_{2}+q_{3})\lambda  \\ 
-q_{1}p_{3}^{2}-2p_{2}p_{3}+q_{1}^{4}-3q_{1}^{2}q_{2}+2q_{1}q_{3}+q_{2}^{2}+c%
\end{array}
& p_{3}\lambda ^{2}+(q_{1}p_{3}+p_{2})\lambda +p_{1}+q_{1}p_{2}+q_{2}p_{3}%
\end{array}%
\right) ,  \label{2.17a}
\end{equation}%
\begin{equation*}
\end{equation*}%
\begin{equation*}
U_{1}=\left( 
\begin{array}{cc}
0 & 1 \\ 
\lambda -2q_{1} & 0%
\end{array}%
\right) ,\ \ \ U_{2}=\left( 
\begin{array}{cc}
-p_{3} & \lambda +q_{1} \\ 
\lambda ^{2}-q_{1}\lambda +q_{1}^{2}-2q_{2} & p_{3}%
\end{array}%
\right) ,
\end{equation*}%
\begin{equation}
\label{2.17b}
\end{equation}%
\begin{equation*}
U_{3}=\left( 
\begin{array}{cc}
-p_{3}\lambda -p_{2}-q_{1}p_{3} & \lambda ^{2}+q_{1}\lambda +q_{2} \\ 
\lambda ^{3}-q_{1}\lambda ^{2}+(q_{1}^{2}-q_{2})\lambda
-p_{3}^{2}-q_{1}^{3}+2q_{1}q_{2}-2q_{3} & p_{3}\lambda +p_{2}+q_{1}p_{3}%
\end{array}%
\right) .
\end{equation*}%
Substituting $q_{1}=\frac{1}{2}\gamma _{1}=\frac{1}{2}u$ we obtain
recursively from (\ref{2.16}) 
\begin{equation*}
(q_{1})_{x}=(X_{1})^{1}\Longleftrightarrow p_{3}=\tfrac{1}{4}u_{x},
\end{equation*}%
\begin{equation*}
(p_{3})_{x}=(X_{1})^{6}\Longleftrightarrow q_{2}=\tfrac{1}{8}u_{xx}+\tfrac{3%
}{8}u^{2}=\tfrac{1}{2}\gamma _{2},
\end{equation*}%
\begin{equation*}
(q_{2})_{x}=(X_{1})^{2}\Longleftrightarrow p_{2}=\tfrac{1}{16}u_{3x}+\tfrac{1%
}{4}u\,u_{x},
\end{equation*}%
\begin{equation*}
(p_{2})_{x}=(X_{1})^{5}\Longleftrightarrow q_{3}=\tfrac{1}{32}u_{4x}+\tfrac{5%
}{16}u\,u_{xx}+\tfrac{5}{32}u_{x}^{2}+\tfrac{5}{16}u^{3}=\tfrac{1}{2}\gamma
_{3},
\end{equation*}%
\begin{equation*}
(q_{3})_{x}=(X_{1})^{3}\Longleftrightarrow p_{1}=\tfrac{1}{64}u_{5x}+\tfrac{9%
}{32}u_{x}u_{xx}+\tfrac{1}{8}u\,u_{3x}+\tfrac{1}{4}u^{2}u_{x},
\end{equation*}%
\begin{equation*}
(p_{1})_{x}=(X_{1})^{4}\Leftrightarrow \tfrac{1}{64}u_{6x}+\tfrac{7}{32}%
u\,u_{4x}+\tfrac{7}{16}u_{x}u_{3x}+\tfrac{21}{64}u_{xx}^{2}+\tfrac{35}{32}%
u^{2}u_{xx}+\tfrac{35}{32}u\,u_{x}^{2}+\tfrac{35}{64}u^{4}+c=\gamma _{4}+c=0,
\end{equation*}%
which coincide with (\ref{2.15}) and the last equation of (\ref{2.13b}).
From evolution equation (\ref{2.17}) we get 
\begin{equation*}
(q_{1})_{t_{2}}=(X_{2})^{1}\Longleftrightarrow u_{t_{2}}=\tfrac{1}{4}u_{xxx}+%
\tfrac{3}{2}u\,u_{x}=\mathcal{K}_{2},
\end{equation*}%
and the remaining equations are differential consequences of $u_{t_{2}}=%
\mathcal{K}_{2}$ and $\gamma _{4}+c=0.$ For example 
\begin{equation*}
(q_{3})_{t_{2}}=(X_{2})^{3}\Longleftrightarrow (\gamma _{4})_{x}=0,\ \ \ \ \
(p_{1})_{t_{2}}=(X_{2})^{4}\Longleftrightarrow (\gamma _{4})_{xx}=0
\end{equation*}%
and so on. Finally, from evolution equation (\ref{2.18}) we get 
\begin{equation*}
(q_{1})_{t_{3}}=(X_{3})^{1}\Longleftrightarrow u_{t_{3}}=\tfrac{1}{16}u_{5x}+%
\tfrac{5}{8}u\,u_{3x}+\tfrac{5}{4}u_{x}u_{xx}+\tfrac{15}{8}u^{2}u_{x}=%
\mathcal{K}_{3},
\end{equation*}%
and again the remaining equations are differential consequences of $%
u_{t_{3}}=\mathcal{K}_{3}$ and $\gamma _{4}+c=0.$Thus, indeed we have the
equivalence between both representations (\ref{2.13a}) and (\ref{2.13b}).
Moreover, under above substitution, St\"{a}ckel Lax matrices $U_{r}$ (\ref%
{2.17b}) turn into KdV matrices $V_{r}~$(\ref{L6})-(\ref{L8}) while $L$
matrix (\ref{2.17a}) is related to the KdV matrix $V_{4}$ under constraint $%
0=\gamma _{4}+c$.
\end{example}

Summarizing results of this section, the $n$-th KdV stationary system (\ref%
{s2}) has the $n$-dimensional St\"{a}ckel representation (\ref{2.9c}),
generated by separation curve (\ref{2.7}).

\section{Hamiltonian representation of the second KdV hierarchy of
stationary systems \label{4}}

Let us briefly collect known facts about the hierarchy of St\"{a}ckel
systems being the Hamiltonian representation of the KdV hierarchy of
stationary systems (\ref{s4}) \cite{blaszak1998,blaszak2008,BS}.

Consider St\"{a}ckel systems generated by the following separation curves on 
$(\lambda ,\mu )$-plane 
\begin{equation}
\lambda ^{2n}+c\lambda ^{-1}+\sum_{r=1}^{n}h_{r}\lambda ^{n-r}=\lambda \mu
^{2},\ \ \ \ \ n\in \mathbb{N}.  \label{4.1}
\end{equation}%
Following the procedure from the previous section we construct $n$
Hamiltonian functions in involution of the form 
\begin{eqnarray}
h_{r} &=&\sum_{i=1}^{n}(-1)^{r+1}\frac{\partial s_{r}}{\partial \lambda _{i}}%
\frac{\lambda _{i}\mu _{i}^{2}}{\Delta _{i}}+\sum_{i=1}^{n}(-1)^{r}\frac{%
\partial s_{r}}{\partial \lambda _{i}}\frac{\lambda _{i}^{2n}+c\lambda
_{i}^{-1}}{\Delta _{i}}  \notag \\
&&  \label{4.3} \\
&=&\mu ^{T}K_{r}G_{1}\mu +V_{r}^{(2n)}+cV_{r}^{(-1)},\quad r=1,\ldots ,n 
\notag
\end{eqnarray}%
where 
\begin{equation}
G_{1}=\text{diag}\left( \frac{\lambda _{1}}{\Delta _{1}},\ldots ,\frac{%
\lambda _{n}}{\Delta _{n}}\right)  \label{4.4}
\end{equation}%
and Killing tensors $K_{r}$ of $G$ takes again the form (\ref{2.9b}). In Vi%
\'{e}te coordinates (\ref{2.11}) 
\begin{equation}
G_{1}^{ij}=%
\begin{cases}
q_{i+j-n}, & i,j=1,\dotsc ,n-1 \\ 
-q_{n}, & i=j=n.%
\end{cases}%
.  \label{4.5}
\end{equation}

Consider the hierarchy of Hamiltonian autonomous dynamical systems, where
each system consists of $n$ evolution equations 
\begin{equation}
\xi _{t_{r}}=X_{r}=\pi dh_{r},\ \ \ \ \ r=1,...,n,\ \ \ \ \ n\in \mathbb{N}
\label{4.6}
\end{equation}%
generated by $n$ Hamiltonian functions $h_{r}$ (\ref{4.3}). The related Lax
isospectral equations are of the form (\ref{4.14}), where $L(\lambda ;\xi )$
matrix takes the form (\ref{2.20}), where now, in Vi\'{e}te coordinates $\xi
=(q,p)$, we have \cite{Blaszak2019a} 
\begin{equation}
u(\lambda ;q)=\lambda ^{n}+\sum_{k=1}^{n}q_{k}\lambda ^{n-k},\ \ \ v(\lambda
;q,p)=-\sum_{k=1}^{n}M_{k}(q,p)\lambda ^{n-k},\ \
M_{k}=\sum_{j=1}^{n}G_{1}^{kj}p_{j}  \label{4.6a}
\end{equation}%
and 
\begin{equation}
w(\lambda ;q,p)=-\lambda \left[ \frac{v(\lambda )^{2}\lambda ^{-1}-\lambda
^{2n}-c\lambda ^{-1}}{u(\lambda )}\right] _{+}.  \label{4.6b}
\end{equation}%
Moreover, matrices $U_{r}(\lambda ;\xi )$ are given by the same formula (\ref%
{2.23}).

Again, the separation curve (\ref{4.1}) is reconstructed from Lax matrix $%
L(\lambda ;\xi )$ through 
\begin{equation*}
\det [L(\lambda ;\xi )-\lambda \mu I]=0.
\end{equation*}

The relation between St\"{a}ckel hierarchy (\ref{4.6}) and the second KdV
stationary hierarchy (\ref{s4}) is as follows.

\begin{theorem}
\cite{BS} For fixed $n\in \mathbb{N}$ and identification $t_{1}$ with $x$ we
get the following equivalence between the St\"{a}ckel system (\ref{4.6}) and
the $n$-th KdV stationary systems (\ref{s4}) 
\begin{equation}
\xi _{t_{r}}=X_{r},\ \ \ \ \ r=1,...,n  \label{4.7}
\end{equation}%
\begin{equation*}
\Updownarrow 
\end{equation*}%
\begin{equation}
\ u_{t_{r}}=\mathcal{K}_{r},\ \ \ \ 0=\tfrac{1}{2}\gamma _{n}(\gamma
_{n})_{xx}-\tfrac{1}{4}[(\gamma _{n})_{x}]^{2}+u\gamma _{n}^{2}+c,\ \ \ \
r=2,...,n,  \label{4.8}
\end{equation}%
where the transformation between jet and Vi\'{e}te coordinates is as follows%
\begin{equation*}
q_{k}=\tfrac{1}{2}\gamma _{k},\ \ \ \ k=2,...,n,\ \ \ \ \ \ p_{k}=\frac{1}{2}%
\sum_{j=1}^{n}(G_{1}^{-1})_{kj}(q_{j})_{x},
\end{equation*}%
with new metric tensor $G_{1}$ (\ref{4.5}). The constraint is encoded in the
last component of the first flow (\ref{4.7}) 
\begin{equation}
0=(p_{n})_{x}-(X_{1})^{2n}\Leftrightarrow 0=\tfrac{1}{2}\gamma _{n}(\gamma
_{n})_{xx}-\tfrac{1}{4}[(\gamma _{n})_{x}]^{2}+u\gamma _{n}^{2}+c.
\label{4.9}
\end{equation}%
Besides, 
\begin{equation}
(q_{1})_{t_{r}}=X_{r}^{1}\Leftrightarrow u_{t_{r}}=\mathcal{K}_{r},\ \ \ \
r=1,...,n  \label{4.9b}
\end{equation}%
and remaining components from systems (\ref{4.7}) for $r=2,...,n$ are
differential consequences of (\ref{4.9}) and (\ref{4.9b}). On the level of
Lax representation of the St\"{a}ckel hierarchy (\ref{4.14}) and the Lax
representation (\ref{s7}) of the KdV stationary system (\ref{s4}), fixing $n$
we have the same identities (\ref{2.15a}) and now 
\begin{equation}
L=V_{n+1}\text{ \ \ \ under constraint \ \ }0=\tfrac{1}{2}\gamma _{n}(\gamma
_{n})_{xx}-\tfrac{1}{4}[(\gamma _{n})_{x}]^{2}+u\gamma _{n}^{2}+c.
\label{4.6c}
\end{equation}
\end{theorem}

Like in the previous case, both systems (\ref{4.7}) and (\ref{4.8}) share
the same solutions.

\begin{example}
\label{e2}Consider once more the case $n=3.$ In $(q,p)$ coordinates, three
Hamiltonians $h_{r}=p^{T}K_{r}Gp+V_{r}^{(6)}+cV_{r}^{(-1)}$, metric tensor $%
G $ and its inverse are 
\begin{align*}
h_{1}&
=(2p_{1}p_{2}+q_{1}p_{2}^{2}-q_{3}p_{3}^{2})+(-q_{1}^{4}+3q_{1}^{2}q_{2}-2q_{1}q_{3}-q_{2}^{2})+cq_{3}^{-1},
\\
h_{2}&
=[p_{1}^{2}+2q_{1}p_{1}p_{2}-2q_{3}p_{2}p_{3}+(q_{1}^{2}-q_{2})p_{2}^{2}-q_{1}q_{3}p_{3}^{2}]+(-q_{1}^{3}q_{2}+q_{1}^{2}q_{3}+2q_{1}q_{2}^{2}-2q_{2}q_{3})+cq_{1}q_{3}^{-1},
\\
h_{3}&
=(-2q_{3}p_{1}p_{3}-q_{3}p_{2}^{2}-2q_{1}q_{3}p_{2}p_{3}-q_{2}q_{3}p_{3}^{2})+(-q_{1}^{3}q_{3}+2q_{1}q_{2}q_{3}-q_{3}^{2})+cq_{2}q_{3}^{-1},
\end{align*}%
\begin{equation}
G_{1}=\left( 
\begin{array}{ccc}
0 & 1 & 0 \\ 
1 & q_{1} & 0 \\ 
0 & 0 & -q_{3}%
\end{array}%
\right) ,\ \ \ \ G_{1}^{-1}=\left( 
\begin{array}{ccc}
-q_{1} & 1 & 0 \\ 
1 & 0 & 0 \\ 
0 & 0 & -\frac{1}{q_{3}}%
\end{array}%
\right) .  \label{4.9a}
\end{equation}%
The related autonomous evolution equations are%
\begin{equation}
\left( 
\begin{array}{c}
q_{1} \\ 
q_{2} \\ 
q_{3} \\ 
p_{1} \\ 
p_{2} \\ 
p_{3}%
\end{array}%
\right) _{x}=X_{1}=\left( 
\begin{array}{l}
\ \ 2p_{2} \\ 
\ \ 2q_{1}p_{2}+2p_{1} \\ 
-2q_{3}p_{3} \\ 
-p_{2}^{2}+4q_{1}^{3}-6q_{1}q_{2}+2q_{3} \\ 
-3q_{1}^{2}+2q_{2} \\ 
\ \ p_{3}^{2}+2q_{1}+cq_{3}^{-2}%
\end{array}%
\right)  \label{4.10}
\end{equation}%
\begin{equation}
\left( 
\begin{array}{c}
q_{1} \\ 
q_{2} \\ 
q_{3} \\ 
p_{1} \\ 
p_{2} \\ 
p_{3}%
\end{array}%
\right) _{t_{2}}=X_{2}=\left( 
\begin{array}{l}
\ \ 2p_{1}+2q_{1}p_{2} \\ 
\ \ 2q_{1}p_{1}-2q_{3}p_{3}+2(q_{1}^{2}-q_{2})p_{2} \\ 
-2q_{1}q_{3}p_{3}-2q_{3}p_{2} \\ 
-2p_{1}p_{2}-2q_{1}p_{2}^{2}+q_{3}p_{3}^{2}+3q_{1}^{2}q_{2}-2q_{1}q_{3}-2q_{2}^{2}-cq_{3}^{-1}
\\ 
\ \ p_{2}^{2}+q_{1}^{3}-4q_{1}q_{2}+2q_{3} \\ 
\ \ 2p_{2}p_{3}+q_{1}p_{3}^{2}-q_{1}^{2}+2q_{2}+cq_{1}q_{3}^{-2}%
\end{array}%
\right)  \label{4.11}
\end{equation}%
\begin{equation}
\left( 
\begin{array}{c}
q_{1} \\ 
q_{2} \\ 
q_{3} \\ 
p_{1} \\ 
p_{2} \\ 
p_{3}%
\end{array}%
\right) _{t_{3}}=X_{3}=\left( 
\begin{array}{l}
-2q_{3}p_{3} \\ 
-2q_{1}q_{3}p_{3}-2q_{3}p_{2} \\ 
-2q_{3}p_{1}-2q_{1}q_{3}p_{2}-2q_{2}q_{3}p_{3} \\ 
\ \ 2q_{3}p_{2}p_{3}+3q_{1}^{2}q_{3}-2q_{2}q_{3} \\ 
\ \ q_{3}p_{3}^{2}-2q_{1}q_{3}-cq_{3}^{-1} \\ 
\ \
2p_{1}p_{3}+p_{2}^{2}+2q_{1}p_{2}p_{3}+q_{2}p_{3}^{2}+q_{1}^{3}-2q_{1}q_{2}+2q_{3}+cq_{2}q_{3}^{-2}%
\end{array}%
\right) .  \label{4.12}
\end{equation}%
Lax representations (\ref{4.14}) of considered equations are as follows 
\begin{equation}
L=\left( 
\begin{array}{cc}
\begin{array}{c}
-p_{2}\lambda ^{2}-(q_{1}p_{2}+p_{1})\lambda +q_{3}p_{3} \\ 
\left. {}\right.%
\end{array}
& 
\begin{array}{c}
\lambda ^{3}+q_{1}\lambda ^{2}+q_{2}\lambda +q_{3} \\ 
\left. {}\right.%
\end{array}
\\ 
\begin{array}{c}
\lambda ^{4}-q_{1}\lambda ^{3}-(q_{2}-q_{1}^{2})\lambda
^{2}-(p_{2}^{2}+q_{1}^{3}-2q_{1}q_{2}+q_{3})\lambda \\ 
-q_{3}p_{3}^{2}+cq_{3}^{-1}%
\end{array}
& p_{2}\lambda ^{2}+(q_{1}p_{2}+p_{1})\lambda -q_{3}p_{3}%
\end{array}%
\right) ,  \label{4.11a}
\end{equation}%
\begin{equation*}
\end{equation*}%
\begin{equation*}
U_{1}=\left( 
\begin{array}{cc}
0 & 1 \\ 
\lambda -2q_{1} & 0%
\end{array}%
\right) ,\ \ \ U_{2}=\left( 
\begin{array}{cc}
-p_{2} & \lambda +q_{1} \\ 
\lambda ^{2}-q_{1}\lambda +q_{1}^{2}-2q_{2} & p_{2}%
\end{array}%
\right) ,
\end{equation*}%
\begin{equation}  \label{4.11b}
\end{equation}%
\begin{equation*}
U_{3}=\left( 
\begin{array}{cc}
-p_{2}\lambda -p_{1}-q_{1}p_{2} & \lambda ^{2}+q_{1}\lambda +q_{2} \\ 
\lambda ^{3}-q_{1}\lambda ^{2}+(q_{1}^{2}-q_{2})\lambda
-p_{2}^{2}-q_{1}^{3}+2q_{1}q_{2}-2q_{3} & p_{2}\lambda +p_{1}+q_{1}p_{2}%
\end{array}%
\right) .
\end{equation*}%
Substituting $q_{1}=\tfrac{1}{2}\gamma _{1}=\tfrac{1}{2}u$ we obtain
recursively from (\ref{4.10}) 
\begin{equation*}
(q_{1})_{x}=(X_{1})^{1}\Longleftrightarrow p_{2}=\tfrac{1}{4}u_{x},
\end{equation*}%
\begin{equation*}
(p_{2})_{x}=(X_{1})^{5}\Longleftrightarrow q_{2}=\tfrac{1}{8}u_{xx}+\tfrac{3%
}{8}u^{2}=\tfrac{1}{2}\gamma _{2},
\end{equation*}%
\begin{equation*}
(q_{2})_{x}=(X_{1})^{2}\Longleftrightarrow p_{1}=\tfrac{1}{16}u_{3x}+\tfrac{1%
}{4}u\,u_{x},
\end{equation*}%
\begin{equation*}
(p_{1})_{x}=(X_{1})^{4}\Longleftrightarrow q_{3}=\tfrac{1}{32}u_{4x}+\tfrac{5%
}{16}u\,u_{xx}+\tfrac{5}{32}u_{x}^{2}+\tfrac{5}{16}u^{3}=\tfrac{1}{2}\gamma
_{3},
\end{equation*}%
\begin{equation*}
(q_{3})_{x}=(X_{1})^{3}\Longleftrightarrow p_{3}=-\tfrac{1}{2}\tfrac{%
(q_{3})_{x}}{q_{3}}=-\tfrac{1}{2}\tfrac{(\gamma _{3})_{x}}{\gamma _{3}},
\end{equation*}%
\begin{equation}
(p_{3})_{x}=(X_{1})^{6}\Longleftrightarrow 0=\tfrac{1}{2}q_{3}(q_{3})_{xx}-%
\tfrac{1}{4}(q_{3}^{2})_{x}+uq_{3}^{2}+c\Rightarrow 0=\pi _{1}\gamma _{3}=%
\mathcal{K}_{4},  \label{p1}
\end{equation}%
which coincide with (\ref{2.15}) with new metric tensor (\ref{4.9a}) and the
last equation of (\ref{4.8}). From evolution equation (\ref{2.17}) we get 
\begin{equation}
(q_{1})_{t_{2}}=(X_{2})^{1}\Longleftrightarrow u_{t_{2}}=\tfrac{1}{4}u_{xxx}+%
\tfrac{3}{2}u\,u_{x}=\mathcal{K}_{2}  \label{p2}
\end{equation}%
and the remaining equations are differential consequences of (\ref{p1}) and (%
\ref{p2}). Finally, from evolution equation (\ref{4.12}) we get 
\begin{equation}
(q_{1})_{t_{3}}=(X_{3})^{1}\Longleftrightarrow u_{t_{3}}=\tfrac{1}{16}u_{5x}+%
\tfrac{5}{8}u\,u_{3x}+\tfrac{5}{4}u_{x}u_{xx}+\tfrac{15}{8}u^{2}u_{x}=%
\mathcal{K}_{3},  \label{p3}
\end{equation}%
and again the remaining equations are differential consequences of (\ref{p1}%
) and (\ref{p3}). Moreover, under above substitution, St\"{a}ckel Lax
matrices $U_{r}$ (\ref{4.11b}) turn into KdV matrices $V_{r}~$(\ref{L6})-(%
\ref{L8}) while $L$ matrix (\ref{4.11a}) is the KdV matrix $V_{4}$ under the
constraint $0=\tfrac{1}{2}\gamma _{3}(\gamma _{3})_{xx}-\tfrac{1}{4}[(\gamma
_{n})_{x}]^{2}+u\gamma _{3}^{2}+c$.Thus, indeed we have the equivalence
between both representations (\ref{4.7}) and (\ref{4.8}), (\ref{4.9}).
\end{example}

Summarizing results of this section, the $n$-th KdV stationary system (\ref%
{s4}) has the $n$-dimensional St\"{a}ckel representation (\ref{4.6}),
generated by the separation curve (\ref{4.1}).

What is interesting, both St\"{a}ckel representations (\ref{2.9c}) and (\ref%
{4.6}) of KdV stationary systems (\ref{2.13b}) and (\ref{4.8}) are related
by a Miura transformation \cite{BS} on the extended phase space (stationary
manifold) $\mathcal{M}_{n}=T^{\ast }Q\times R$ $\ni (q,p,c)$ and in
consequence both systems have bi-Hamiltonian representation on $\mathcal{M}%
_{n}$ \cite{blaszak1998}.

\section{The Painlev\'{e} deformation of St\"{a}ckel systems \label{5}}

In the sequence of papers \cite{Blaszak2021,BlaszakX1,BlaszakX2} we have
presented a systematic deformation of an autonomous St\"{a}ckel systems
generated by separation curves of the form 
\begin{equation}
\sum_{k=-m}^{2n+2-m}\lambda ^{k}+\sum_{r=1}^{n}h_{r}\lambda ^{n-r}=\lambda
^{m}\mu ^{2},\ \ \ \ \ \ \ m\in \{0,1,...,n+1\}  \label{5.0}
\end{equation}%
to non-autonomous Painlev\'{e}-type systems 
\begin{equation}
\frac{d\xi }{dt_{r}}=Y_{r}(\xi ,t)=\pi dH_{r}(\xi ,t),\quad r=1,\ldots ,n.
\label{5.1}
\end{equation}%
In consequence new Hamiltonian functions $H_{r}$ fulfill the Frobenius
integrability conditions%
\begin{equation}
\{H_{r},H_{s}\}+\frac{\partial H_{r}}{\partial t_{s}}-\frac{\partial H_{s}}{%
\partial t_{r}}=f_{rs}(t_{1},\ldots ,t_{n}),\quad r,s=1,\dots ,n,
\label{5.2}
\end{equation}%
where $f_{rs}$ are some functions not depending on the phase-space variables 
$\xi $, only on the parameters $t_{j}$, and related Hamiltonian vector
fields $Y_{k}(\xi ,t)$ satisfy 
\begin{equation}
\left[ Y_{s},Y_{r}\right] +\frac{\partial Y_{r}}{\partial t_{s}}-\frac{%
\partial Y_{s}}{\partial t_{r}}=0,\quad r,s=1,\dots ,n,  \label{5.3}
\end{equation}%
as $\pi d\{H_{r},H_{s}\}=-\left[ Y_{r},Y_{s}\right] $. Therefore, the set of
non-autonomous evolution equations (\ref{5.1}) has again common (at least
local) solutions $\xi (t_{1},\ldots ,t_{n},\xi _{0})$ through each point $%
\xi _{0}$ of $M$ \cite{Fecko,Iwasaki,Lundell}.

Observe, that it is always possible to choose another Hamiltonians $%
H_{r}\rightarrow H_{r}+\varphi _{r}(t_{1},...,t_{n})$, defining the same
dynamical systems (\ref{5.1}), which satisfy 
\begin{equation}
\{H_{r},H_{s}\}+\frac{\partial H_{r}}{\partial t_{s}}-\frac{\partial H_{s}}{%
\partial t_{r}}=0,\quad r,s=1,\dots ,n.  \label{5.2a}
\end{equation}

Besides, the non-autonomous Hamiltonian equations (\ref{5.1}) are
represented by the so-called isomonodromic Lax representation%
\begin{equation}
\frac{d}{dt_{k}}L(\lambda ;\xi ,t)=[U_{k}(\lambda ;\xi ,t),L(\lambda ;\xi
,t)]+\lambda ^{m}\frac{\partial U_{k}(\lambda ;\xi ,t)}{\partial \lambda },
\label{5.5}
\end{equation}%
being the compatibility condition for a system of linear equations 
\begin{equation}
\lambda ^{m}\frac{\partial \Psi }{\partial \lambda }=L(\lambda ;\xi ,t)\Psi
,\qquad \frac{d}{dt_{k}}\Psi =U_{k}(\lambda ;\xi ,t)\Psi .  \label{5.4}
\end{equation}

The deformation procedure for St\"{a}ckel systems considered in previous
sections, is as follows \cite{BlaszakX1}. First, we deform geodesic
constants of motion $E_{r}=\mu ^{T}K_{r}G\mu $ by terms linear in momenta 
\begin{equation}
\mathcal{E}_{r}=\ E_{r}+W_{r}=\mu ^{T}K_{r}G\mu +\mu ^{T}J_{r},\quad
r=2,\dotsc ,n,\ \ \ \ W_{1}=0,  \label{5.6}
\end{equation}%
generated by Killing vectors $J_{r}$ of metric tensors $G_{0}$ (\ref{2.11aa}%
) and $G_{1}$ (\ref{4.5}). Actually, in $(q,p)$ coordinates, \quad 
\begin{equation}
W_{r}=\sum\limits_{k=n-r+2}^{n}(n+1-k)q_{r+k-n-2}\,p_{k},  \label{5.7}
\end{equation}%
for metric $G_{0}$\ and 
\begin{equation}
W_{r}=\sum\limits_{k=n-r+1}^{n-1}(n-k)q_{r+k-n-1}\,p_{k},  \label{5.8}
\end{equation}%
for metric $G_{1}$, respectively.

The Hamiltonians $\mathcal{E}_{r}$ in (\ref{5.6}) span a Lie algebra $%
\mathfrak{g}=\mathrm{span}\{\mathcal{E}_{r}\in C^{\infty }(M)\colon $ $%
r=1,\ldots ,n\}$ with the following commutation relations 
\begin{equation}
\{\mathcal{E}_{1},\mathcal{E}_{r}\}=0,\quad r=2,\dots ,n,  \label{5.10}
\end{equation}%
where 
\begin{equation}
\{\mathcal{E}_{r},\mathcal{E}_{s}\}=(s-r)\mathcal{E}_{r+s-n-2},\ \ \ \
r,s=2,...,n,  \label{5.11}
\end{equation}%
in the first case and 
\begin{equation}
\{\mathcal{E}_{r},\mathcal{E}_{s}\}=(s-r)\mathcal{E}_{r+s-n-1},\ \ \ \
r,s=2,...,n,  \label{5.12}
\end{equation}%
in the second case. We use the convention that $\mathcal{E}_{r}=0$ as soon
as $r\leq 0$ or $r>n$. The algebra $\mathfrak{g}$ has an Abelian subalgebra 
\begin{equation}
\mathfrak{a}=\mathrm{span}\left\{ \mathcal{E}_{1},\dotsc ,\mathcal{E}%
_{\kappa }\right\} ,\ \ \ \ \ \ \kappa =\left[ \frac{n+3}{2}\right] \text{
and }\kappa =\left[ \frac{n+2}{2}\right] \text{ respectively}.  \label{5.13}
\end{equation}

As the Hamiltonians $\mathcal{E}_{r}$ in (\ref{5.6}) do not commute, they do
not constitute a Liouville integrable system. In particular, there is no
reason to expect that they will possess a common, multi-time solution for a
given initial data $\xi _{0}$. However, in \cite{Blaszak2021} we found
polynomial-in-times deformations $H_{r}(t_{1},\ldots ,t_{n})$ of the
Hamiltonians $\mathcal{E}_{r}$ such that the Hamiltonians $H_{r}$ satisfy
the Frobenius integrability condition (\ref{5.2}). More specifically, the
deformed Hamiltonians $H_{r}$ are given by 
\begin{align}
H_{r}& =\mathcal{E}_{r},\quad \text{for $r=1,\dotsc ,\kappa $},  \notag \\
H_{r}& =\sum_{j=1}^{r}\zeta _{r,j}(t_{1},\dotsc ,t_{r-1})\mathcal{E}%
_{j},\quad \zeta _{r,r}=1,\quad \text{for $r=\kappa +1,\dotsc ,n.$}
\label{5.14}
\end{align}%
where $\zeta _{r,j}(t)$ are some polynomial functions of evolution
parameters, determined uniquely from Frobenius conditions (\ref{5.2a}).

In the second step of deformation process we include the appropriate
potentials from (\ref{2.9}) and (\ref{4.3}). In order to preserve Frobenius
conditions, we have supplement these potentials by extra lower order
potentials with time dependent coefficients. Actually, the deformation of
Hamiltonians (\ref{2.9}) takes the form 
\begin{equation}
h_{r}=E_{r}+V_{r}^{(2n+1)}+cV_{r}^{(n)}\longrightarrow
h_{r}^{W}=h_{r}+W_{r}+\sum_{k=n}^{2n-1}c_{k}(t)V_{r}^{(k)},\quad r=1,\ldots
,n  \label{5.16}
\end{equation}%
and Hamiltonians (\ref{4.3}) the respective form 
\begin{equation}
h_{r}=E_{r}+V_{r}^{(2n)}+cV_{r}^{(-1)}\longrightarrow
h_{r}^{W}=h_{r}+W_{r}+\sum_{k=n}^{2n-1}d_{k}(t)V_{r}^{(k)},\quad r=1,\ldots
,n.  \label{5.17}
\end{equation}%
where $c_{k}(t)$ and $d_{k}(t)$ are again some polynomial functions of
evolution parameters, determined uniquely from Frobenius conditions for
functions $H_{r}$ 
\begin{align}
H_{r}& =h_{r}^{W},\quad \text{for $r=1,\dotsc ,\kappa $},  \notag \\
H_{r}& =\sum_{j=1}^{r}\zeta _{r,j}(t_{1},\dotsc ,t_{r-1})h_{j}^{W},\quad
\zeta _{r,r}=1,\quad \text{for $r=\kappa +1,\dotsc ,n.$}  \label{5.18}
\end{align}%
with the same $\zeta _{r,j}(t)$ coefficients as in (\ref{5.14}). The details
of the deformation procedure as well as an appropriate values of
coefficients $\zeta _{r,j}(t)$, $c_{k}(t)$ and $d_{k}(t)$ the reader can
find in \cite{BlaszakX1}.

Both hierarchies of non-autonomous Frobenius integrable systems have
isomonodromic Lax representations \cite{BlaszakX2}, so are represented by a
Painlev\'{e}-type equations.

\begin{theorem}
\cite{BlaszakX2} For the first hierarchy, generated by Hamiltonians (\ref%
{5.18}) and (\ref{5.16}), isomonodromic Lax representations are of the form 
\begin{equation}
\frac{d}{dt_{k}}L(\lambda ;\xi ,t)=[U_{k}(\lambda ;\xi ,t),L(\lambda ;\xi
,t)]+\frac{\partial U_{k}(\lambda ;\xi ,t)}{\partial \lambda },\ \ \ \ \
k=1,...,n.  \label{5.18.2}
\end{equation}%
The $L(\lambda ;\xi ,t)$ matrix takes the form (\ref{2.20}), (\ref{2.21})
where now 
\begin{equation}
w(\lambda ;q,p,t)=-\left[ \frac{v(\lambda )^{2}-\lambda
^{2n+1}-\sum_{k=n}^{2n-1}c_{k}(t)\lambda ^{k}}{u(\lambda )}\right] _{+}.
\label{5.18.3}
\end{equation}%
Moreover, 
\begin{align}
U_{r}(\lambda ;\xi ,t)& =\left[ \frac{u_{r}(\lambda )L(\lambda )}{u(\lambda )%
}\right] _{+},\quad \text{for $r=1,\dotsc ,\kappa $},  \notag \\
U_{r}(\lambda ;\xi ,t)& =\sum_{j=1}^{r}\zeta _{r,j}(t_{1},\dotsc ,t_{r-1}) 
\left[ \frac{u_{j}(\lambda )L(\lambda )}{u(\lambda )}\right] _{+},\quad
\zeta _{r,r}=1,\quad \text{for $r=\kappa +1,\dotsc ,n.$}  \label{5.18.4}
\end{align}%
For the second hierarchy, generated by Hamiltonians (\ref{5.18}) and (\ref%
{5.17}), isomonodromic Lax representations are of the form 
\begin{equation}
\frac{d}{dt_{k}}L(\lambda ;\xi ,t)=[U_{k}(\lambda ;\xi ,t),L(\lambda ;\xi
,t)]+\lambda \frac{\partial U_{k}(\lambda ;\xi ,t)}{\partial \lambda }.
\label{5.18.6}
\end{equation}%
The $L(\lambda ;\xi ,t)$ matrix takes the form (\ref{2.20}) with entries $%
u(\lambda ;q)$ and $v(\lambda ;q,p)$ given by (\ref{4.6a}), where now 
\begin{equation}
w(\lambda ;q,p,t)=-\lambda \left[ \frac{v(\lambda )^{2}\lambda ^{-1}-\lambda
^{2n}-\sum_{k=n}^{2n-1}c_{k}(t)\lambda ^{k}-c\lambda ^{-1}}{u(\lambda )}%
\right] _{+}.  \label{5.18.8}
\end{equation}%
$U_{r}(\lambda ;\xi ,t)$ matrices are again of the form (\ref{5.18.4}). \ 
\end{theorem}

The first members of the hierarchy (\ref{5.16}), (\ref{5.18}) are determined
by the following Hamiltonians%
\begin{equation}
\begin{array}{l}
n=1:\ \ \ h_{1}^{W}=E_{1}+V_{1}^{(3)}+(t_{1}+c)V_{1}^{(1)},\ \ \ \
H_{1}=h_{1}^{W}, \\ 
\\ 
n=2:\ \ \ h_{r}^{W}=\mathcal{E}%
_{r}+V_{r}^{(5)}+3t_{2}V_{r}^{(4)}+(t_{1}+c)V_{r}^{(3)},\ \ \ \
H_{r}=h_{r}^{W},\ \ \ \ r=1,2, \\ 
\\ 
n=3:\ \ \ h_{r}^{W}=\mathcal{E}%
_{r}+V_{r}^{(7)}+5t_{3}V_{r}^{(5)}+3t_{2}V_{r}^{(4)}+(t_{1}+\frac{15}{2}%
t_{3}^{2}+c)V_{r}^{(3)},\ \ \ \ H_{r}=h_{r}^{W},\ \ \ \ r=1,2,3, \\ 
\\ 
n=4:\ \ \ h_{r}^{W}=\mathcal{E}%
_{r}+V_{r}^{(9)}+7t_{4}V_{r}^{(7)}+5t_{3}V_{r}^{(6)}+(3t_{2}+\frac{35}{2}%
t_{4}^{2})V_{r}^{(5)}+(t_{1}+21t_{3}t_{4}+c)V_{r}^{(4)}, \\ 
\\ 
\ \ \ \ \ \ \ \ \ \ \ \ \ H_{r}=h_{r}^{W},\ \ \ \
r=1,2,3,~~~H_{4}=h_{4}^{W}+t_{3}h_{1}^{W}, \\ 
\\ 
n=5:\ \ \ h_{r}^{W}=\mathcal{E}%
_{r}+V_{r}^{(11)}+9t_{5}V_{r}^{(9)}+7t_{4}V_{r}^{(8)}+5t_{3}V_{r}^{(7)}+(3t_{2}+45t_{4}t_{5})V_{r}^{(6)}
\\ 
\\ 
\ \ \ \ \ \ \ \ \ \ \ \ \ \ \ \ \ \ \ \ \ \ \ \ \
+(t_{1}+27t_{3}t_{5}+14t_{4}^{2}+\frac{105}{2}t_{5}^{3}+c)V_{r}^{(5)}, \\ 
\\ 
\ \ \ \ \ \ \ \ \ \ \ \ \ H_{r}=h_{r}^{W},\ \ \ \
r=1,...,4,~~~H_{5}=h_{5}^{W}+t_{4}h_{2}^{W}+2t_{3}h_{1}^{W}, \\ 
\vdots%
\end{array}
\label{5.18a}
\end{equation}%
while the first members of the second hierarchy (\ref{5.17}), (\ref{5.18})
are determined by Hamiltonians%
\begin{equation}
\begin{array}{l}
n=1:\ \ \ h_{1}^{W}=E_{1}+V_{1}^{(2)}+t_{1}V_{1}^{(1)}+cV_{1}^{(-1)},\ \ \ \
H_{1}=h_{1}^{W}, \\ 
\\ 
n=2:\ \ \ h_{r}^{W}=\mathcal{E}%
_{r}+V_{r}^{(4)}+3t_{2}V_{r}^{(3)}+(t_{1}+3t_{2}^{2})V_{r}^{(2)}+cV_{r}^{(-1)},\ \ \ \ H_{r}=h_{r}^{W},\ \ \ \ r=1,2,
\\ 
\\ 
n=3:\ \ \ h_{r}^{W}=\mathcal{E}%
_{r}+V_{r}^{(6)}+5t_{3}V_{r}^{(5)}+(3t_{2}+10t_{3}^{2})V_{r}^{(4)}+(t_{1}+10t_{2}t_{3}+10t_{3}^{3})V_{r}^{(3)}+cV_{r}^{(-1)},
\\ 
\\ 
\ \ \ \ \ \ \ \ \ \ \ \ \ \ H_{r}=h_{r}^{W},\ \ \ \ r=1,2,\ \ \
H_{3}=h_{3}^{W}+t_{2}h_{1}^{W}, \\ 
\\ 
n=4:\ \ \ h_{r}^{W}=\mathcal{E}%
_{r}+V_{r}^{(8)}+7t_{4}V_{r}^{(7)}+(5t_{3}+21t_{4}^{2})V_{r}^{(6)}+(3t_{2}+28t_{3}t_{4}+35t_{4}^{2})V_{r}^{(5)}
\\ 
\\ 
\ \ \ \ \ \ \ \ \ \ \ \ \ \ \ \ \ \ \ \ \ \ \ \ +(t_{1}+14t_{2}t_{4}+\frac{15%
}{2}t_{3}^{2}+63t_{3}t_{4}^{2}+35t_{4}^{4})V_{r}^{(4)}+cV_{r}^{(-1)}, \\ 
\\ 
\ \ \ \ \ \ \ \ \ \ \ \ \ \ H_{r}=h_{r}^{W},\ \ \ \ r=1,2,3,\ \ \
H_{4}=h_{4}^{W}+t_{3}h_{2}^{W}+2t_{2}h_{1}^{W}, \\ 
\\ 
n=5:\ \ \ h_{r}^{W}=\mathcal{E}%
_{r}+V_{r}^{(10)}+9t_{5}V_{r}^{(9)}+(7t_{4}+36t_{5}^{2})V_{r}^{(8)}+(5t_{3}+54t_{4}t_{5}+84t_{5}^{3})V_{r}^{(7)}
\\ 
\\ 
\ \ \ \ \ \ \ \ \ \ \ \ \ \ \ \ \ \ \ \ \ \ +(3t_{2}+36t_{3}t_{5}+\frac{35}{2%
}t_{4}^{2}+180t_{4}t_{5}^{2}+126t_{5}^{4})V_{r}^{(6)} \\ 
\\ 
\ \ \ \ \ \ \ \ \ \ \ \ \ \ \ \ \ \ \ \ \ \
+(t_{1}+21t_{3}t_{4}+18t_{2}t_{5}+108t_{4}^{2}t_{5}+108t_{3}t_{5}^{2}+336t_{4}t_{5}^{3}+126t_{5}^{5})V_{r}^{(5)}+cV_{r}^{(-1)},
\\ 
\\ 
\ \ \ \ \ \ \ \ \ \ \ \ \ \ H_{r}=h_{r}^{W},\ \ r=1,2,3,\ \
H_{4}=h_{4}^{W}+t_{3}h_{1}^{W},\ \
H_{5}=h_{5}^{W}+t_{4}h_{3}^{W}+2t_{3}h_{2}^{W}+(3t_{2}-\frac{1}{2}%
t_{4}^{2})h_{1}^{W}, \\ 
\vdots%
\end{array}
\label{5.18b}
\end{equation}

The first hierarchy (\ref{5.18a}) is the complete hierarchy of Painlev\'{e}
I ($P_{I}$) systems, as its first, one dimensional, system is a famous $%
P_{I} $ equation. Indeed, denoted $q_{1}=q,\ p_{1}=p$ and $t_{1}=t$ we have 
\begin{equation*}
H=h_{1}^{W}=p^{2}+q^{3}+(t+c)q
\end{equation*}%
\begin{equation*}
\Downarrow
\end{equation*}%
\begin{equation*}
q_{t}=2p,\ \ \ p_{t}=-3q^{2}-t-c
\end{equation*}%
\begin{equation*}
\Updownarrow
\end{equation*}%
\begin{equation}
q_{tt}+6q^{2}+2(t+c)=0  \label{5.19}
\end{equation}%
The isomonodromic Lax representation (\ref{5.5}) is as follows 
\begin{equation}
L=\left( 
\begin{array}{cc}
-p & \lambda +q \\ 
\lambda ^{2}-q\lambda +q^{2}+t+c & p%
\end{array}%
\right) ,\ \ \ \ \ U=\left( 
\begin{array}{cc}
0 & 1 \\ 
\lambda -2q & 0%
\end{array}%
\right) .  \label{5.20}
\end{equation}%
On the other hand, equation (\ref{5.19}) is equivalent to $P_{I}$ equation 
\begin{equation}
q_{tt}=3q^{2}+t  \label{5.21}
\end{equation}%
after rescaling $q\rightarrow -2^{\frac{1}{5}}q$, $t\rightarrow 2^{-\frac{2}{%
5}}t-c$.

The second, two dimensional system, from that hierarchy is as follows%
\begin{align*}
H_{1}&
=h_{1}^{W}=2p_{1}p_{2}+q_{1}p_{2}^{2}-q_{1}^{4}+3q_{1}^{2}q_{2}-q_{2}^{2}+3t_{2}(q_{2}-q_{1}^{2})+(t_{1}+c)q_{1},
\\
H_{2}&
=h_{2}^{W}=p_{1}^{2}+2q_{1}p_{1}p_{2}+(q_{1}^{2}-q_{2})p_{2}^{2}+p_{2}-q_{1}^{3}q_{2}+2q_{1}q_{2}^{2}-3t_{2}q_{1}q_{2}+(t_{1}+c)q_{2},
\\
&
\end{align*}%
\begin{equation*}
\{H_{1},H_{2}\}+\frac{\partial H_{1}}{\partial t_{2}}-\frac{\partial H_{2}}{%
\partial t_{1}}=3t_{2}.
\end{equation*}%
Thus, 
\begin{equation*}
\left( 
\begin{array}{l}
q_{1} \\ 
q_{2} \\ 
p_{1} \\ 
p_{2}%
\end{array}%
\right) _{t_{1}}=\left( 
\begin{array}{l}
\ \ 2p_{2} \\ 
\ \ 2p_{1}+2q_{1}p_{2} \\ 
-p_{2}^{2}+4q_{1}^{3}-6q_{1}q_{2}+6t_{2}q_{1}-t_{1}-c \\ 
-3q_{1}^{2}+2q_{2}-3t_{2}%
\end{array}%
\right) =Y_{1}
\end{equation*}%
\begin{equation*}
\Updownarrow
\end{equation*}%
\begin{equation}
(q_{1})_{t_{1}t_{1}}=-6q_{1}^{2}+4q_{2}-6t_{2},\ \ \ \ (q_{2})_{t_{1}t_{1}}=%
\frac{1}{2}%
[(q_{1})_{t_{1}}]^{2}+2q_{1}^{3}-8q_{1}q_{2}+6t_{2}q_{1}-2t_{1}-2c,
\label{5.21a}
\end{equation}%
\begin{equation*}
\end{equation*}%
\begin{equation*}
\left( 
\begin{array}{l}
q_{1} \\ 
q_{2} \\ 
p_{1} \\ 
p_{2}%
\end{array}%
\right) _{t_{2}}=\left( 
\begin{array}{l}
\ \ 2q_{1}p_{2}+2p_{1} \\ 
\ \ 2q_{1}p_{1}+2(q_{1}^{2}-q_{2})p_{2}+1 \\ 
-2q_{1}p_{2}^{2}-2p_{1}p_{2}+3q_{1}^{2}q_{2}-2q_{2}^{2}+3t_{2}q_{2} \\ 
\ \ p_{2}^{2}+q_{1}^{3}-4q_{1}q_{2}+3t_{2}q_{1}-t_{1}-c%
\end{array}%
\right) =Y_{2}
\end{equation*}%
and 
\begin{equation*}
\lbrack Y_{2},Y_{1}]+\frac{\partial Y_{1}}{\partial t_{2}}-\frac{\partial
Y_{2}}{\partial t_{1}}=0.
\end{equation*}%
Notice that eliminating $q_{2}$ from (\ref{5.21a}) we get the forth order
equation for $q\equiv q_{1}$ 
\begin{equation*}
\tfrac{1}{4}q_{t_{1}t_{1}t_{1}t_{1}}+5qq_{t_{1}t_{1}}+\tfrac{5}{2}%
(q_{t_{1}})^{2}+10q^{3}+6t_{2}q+2t_{1}+2c=0,
\end{equation*}%
which is the second equation from the standard $P_{I}$ hierarchy.

The isomonodromic Lax representation (\ref{5.5}) is of the form%
\begin{equation*}
L=\left( 
\begin{array}{cc}
-p_{2}\lambda -p_{1}-q_{1}p_{2} & \lambda ^{2}+q_{1}\lambda +q_{2} \\ 
\lambda ^{3}-q_{1}\lambda ^{2}+(q_{1}^{2}-q_{2}+3t_{2})\lambda
-p_{2}^{2}+2q_{1}q_{2}-3t_{2}q_{1}+t_{1}+c & p_{2}\lambda +p_{1}+q_{1}p_{2}%
\end{array}%
\right) ,
\end{equation*}%
\begin{equation*}
U_{1}=\left( 
\begin{array}{cc}
0 & 1 \\ 
\lambda -2q_{1} & 0%
\end{array}%
\right) ,\ \ \ \ U_{2}=\left( 
\begin{array}{cc}
-p_{2} & \lambda +q_{1} \\ 
\lambda ^{2}-q_{1}\lambda +q_{1}^{2}-2q_{2}+3t_{2} & p_{2}%
\end{array}%
\right) .
\end{equation*}

The third, three dimensional system, from that hierarchy (\ref{5.18a}), i.e.
the deformed system from Example \ref{e1}, is generated by Hamiltonians%
\begin{align}
H_{1}& =h_{1}^{W}=h_{1}+5t_{3}V_{1}^{(5)}+3t_{2}V_{1}^{(4)}+(t_{1}+\tfrac{15%
}{2}t_{3}^{2})V_{1}^{(3)},  \notag \\
H_{2}& =h_{2}^{W}=h_{2}+p_{3}+5t_{3}V_{2}^{(5)}+3t_{2}V_{2}^{(4)}+(t_{1}+%
\tfrac{15}{2}t_{3}^{2})V_{2}^{(3)},  \label{n3} \\
H_{3}&
=h_{3}^{W}=h_{3}+2p_{2}+q_{1}p_{3}+5t_{3}V_{3}^{(5)}+3t_{2}V_{3}^{(4)}+(t_{1}+%
\tfrac{15}{2}t_{3}^{2})V_{3}^{(3)},  \notag
\end{align}%
where functions $h_{1},h_{2}$ and $h_{3}$ are given in Example \ref{e1} and 
\begin{equation*}
V_{1}^{(4)}=q_{2}-q_{1}^{2},\ \ \ V_{2}^{(4)}=q_{3}-q_{1}q_{2},\ \ \
V_{3}^{(4)}=-q_{1}q_{3},
\end{equation*}%
\begin{equation*}
V_{1}^{(5)}=q_{1}^{3}-2q_{1}q_{2}+q_{3},\ \ \
V_{2}^{(5)}=q_{1}^{2}q_{2}-q_{1}q_{3}-q_{2}^{2},\ \ \
V_{3}^{(5)}=q_{1}^{2}q_{3}-q_{2}q_{3}.
\end{equation*}%
The related Hamiltonian vector fields $Y_{r}=\pi dH_{r}$, $r=1,2,3$ fulfill
Frobenius conditions (\ref{5.3}) and evolution equations $\xi _{t_{r}}=Y_{r}$
have the following isomonodromic Lax representations%
\begin{equation*}
L=\left( 
\begin{array}{cc}
-p_{3}\lambda ^{2}-(p_{2}+q_{1}p_{3})\lambda -(p_{1}+q_{1}p_{2}+q_{2}p_{3})
& \lambda ^{3}+q_{1}\lambda ^{2}+q_{2}\lambda +q_{3} \\ 
L_{21} & -L_{11}%
\end{array}%
\right) ,
\end{equation*}%
\begin{equation*}
L_{21}=\lambda ^{4}-q_{1}\lambda ^{3}-(V_{1}^{(4)}-5t_{3})\lambda
^{2}-(V_{1}^{(5)}+p_{3}^{2}-5q_{1}t_{3}-3t_{2})\lambda
-V_{1}^{(6)}-q_{1}p_{3}^{2}-2p_{2}p_{3}-5t_{3}V_{1}^{(4)}-3t_{2}q_{1}+t_{1}+%
\tfrac{15}{2}t_{3}^{2}+c,
\end{equation*}%
\begin{equation*}
V_{1}^{(6)}=-q_{1}^{4}+3q_{1}^{2}q_{2}-2q_{1}q_{3}-q_{2}^{2}.
\end{equation*}%
\begin{equation*}
\end{equation*}%
\begin{equation*}
U_{1}=\left( 
\begin{array}{cc}
0 & 1 \\ 
\lambda -2q_{1} & 0%
\end{array}%
\right) ,\ \ \ U_{2}=\left( 
\begin{array}{cc}
-p_{3} & \lambda +q_{1} \\ 
\lambda ^{2}-q_{1}\lambda +q_{1}^{2}-2q_{2}+5t_{3} & p_{3}%
\end{array}%
\right) ,
\end{equation*}%
\begin{equation*}
U_{3}=\left( 
\begin{array}{cc}
-p_{3}\lambda -q_{1}p_{3}-p_{2} & \lambda ^{2}+q_{1}\lambda +q_{2} \\ 
\lambda ^{3}-q_{1}\lambda ^{2}-(V_{1}^{(4)}-5t_{3})\lambda
-V_{1}^{(5)}-q_{3}-p_{3}^{2}-5q_{1}t_{3}+3t_{2} & p_{3}\lambda
+q_{1}p_{3}+p_{2}%
\end{array}%
\right)
\end{equation*}

The explicit form of isomonodromic Lax representations of higher dimensional
systems from the $P_{I}$ hierarchy (\ref{5.18a}) can be constructed with the
help of (\ref{5.18.3}) and (\ref{5.18.4}).

The first attempt to the hierarchy of $P_{I}$ systems was done in \cite%
{Takasaki} by Takasaki, where the author started from the opposite side,
i.e. from string equations of KP (KdV in particular). Using such formalism
he was able to construct, for each $n$, only the first Painlev\'{e} equation 
\begin{equation*}
\frac{d\xi }{dt_{1}}=Y_{1}(\xi ,t)=\pi dH_{1}(\xi ,t),
\end{equation*}%
from the system (\ref{5.1}), together with its isomonodromic Lax
representation. He failed in construction the remaining equations from the $%
P_{I}$ system (\ref{5.1}), with $k=2,...,n,$ as he did not control the
perturbation terms $W_{k}$ (\ref{5.7}) in remaining Hamiltonians. Now we
know that they are generated by Killing vectors (\ref{5.7})\ of the metric
tensor $G_{0}$.

Now, let us turn to the second hierarchy of Painlev\'{e} systems (\ref{5.18b}%
). The first, one dimensional, system is as follows:%
\begin{equation*}
H=h_{1}^{W}=-qp^{2}-q^{2}+tq+cq^{-1}
\end{equation*}%
\begin{equation*}
\Downarrow
\end{equation*}%
\begin{equation*}
q_{t}=-2qp,\ \ \ \ p_{t}=p^{2}+2q-t+cq^{-2}
\end{equation*}%
\begin{equation*}
\Updownarrow
\end{equation*}%
\begin{equation}
qq_{tt}=\frac{1}{2}q_{t}^{2}-4q^{3}+2tq^{2}-2c.  \label{5.22}
\end{equation}

The isomonodromic Lax representation takes the form (\ref{5.18.6}), where 
\begin{equation*}
L=\left( 
\begin{array}{cc}
qp & \lambda +q \\ 
\lambda ^{2}+(-q+t)\lambda +qp^{2}+cq^{-1} & -qp%
\end{array}%
\right) ,\ \ \ \ U=\left( 
\begin{array}{cc}
0 & 1 \\ 
\lambda -2q+t & 0%
\end{array}%
\right) .
\end{equation*}%
One can verify that equation (\ref{5.22}) is the thirty-fourth Painlev\'{e} (%
$P_{34}$) equation from the Gambier list.

The second, two dimensional system from the hierarchy, is generated by
Hamiltonians%
\begin{align*}
H_{1}&
=h_{1}^{W}=p_{1}^{2}-q_{2}p_{2}^{2}+q_{1}^{3}-2q_{1}q_{2}+3t_{2}(q_{2}-q_{1}^{2})+(3t_{2}^{2}+t_{1})q_{1}+cq_{2}^{-1},
\\
H_{2}&
=h_{2}^{W}=-2q_{2}p_{1}p_{2}-q_{1}q_{2}p_{2}^{2}+p_{1}+q_{1}^{2}q_{2}-q_{2}^{2}-3t_{2}q_{1}q_{2}+(3t_{2}^{2}+t_{1})q_{2}+cq_{1}q_{2}^{-1},
\\
&
\end{align*}%
\begin{equation*}
\{H_{1},H_{2}\}+\frac{\partial H_{1}}{\partial t_{2}}-\frac{\partial H_{2}}{%
\partial t_{1}}=t_{1}+3t_{2}^{2}.
\end{equation*}%
Thus, 
\begin{equation*}
\left( 
\begin{array}{l}
q_{1} \\ 
q_{2} \\ 
p_{1} \\ 
p_{2}%
\end{array}%
\right) _{t_{1}}=\left( 
\begin{array}{l}
\ \ 2p_{1} \\ 
-2q_{2}p_{2} \\ 
-3q_{1}^{2}+2q_{2}+6t_{2}q_{1}-3t_{2}^{2}-t_{1} \\ 
\ \ p_{2}^{2}+2q_{1}-3t_{2}+cq_{2}^{-2}%
\end{array}%
\right) =Y_{1}
\end{equation*}%
\begin{equation*}
\Updownarrow
\end{equation*}%
\begin{equation}
(q_{1})_{t_{1}t_{1}}=-6q_{1}^{2}+4q_{2}+12t_{2}q_{1}-6t_{2}^{2}-2t_{1},\ \ \
\ \ q_{2}(q_{2})_{t_{1}t_{1}}=\frac{1}{2}%
[(q_{2})_{t_{1}}]^{2}-4q_{1}q_{2}^{2}+6t_{2}q_{2}^{2}-2c,  \label{5.22a}
\end{equation}%
\begin{equation*}
\end{equation*}%
\begin{equation*}
\left( 
\begin{array}{l}
q_{1} \\ 
q_{2} \\ 
p_{1} \\ 
p_{2}%
\end{array}%
\right) _{t_{2}}=\left( 
\begin{array}{l}
-2q_{2}p_{2}+1 \\ 
-2q_{2}p_{1}-2q_{1}q_{2}p_{2} \\ 
\ \ q_{2}p_{2}^{2}-2q_{1}q_{2}+3t_{2}q_{2}-cq_{2}^{-1} \\ 
\ \
2p_{1}p_{2}+q_{1}p_{2}^{2}-q_{1}^{2}+2q_{2}+3t_{2}q_{1}-3t_{2}^{2}-t_{1}+cq_{1}q_{2}^{-2}%
\end{array}%
\right) =Y_{2}
\end{equation*}%
Notice that eliminating $q_{2}$ from (\ref{5.22a}) we get the forth order
equation for $q\equiv q_{1}$%
\begin{eqnarray*}
0 &=&-\tfrac{1}{32}q_{t_{1}t_{1}}q_{t_{1}t_{1}t_{1}t_{1}}+\tfrac{1}{64}%
(q_{t_{1}t_{1}t_{1}})^{2}-\tfrac{1}{16}%
(3q^{2}-6t_{2}q+3t_{2}^{2}+t_{1})q_{t_{1}t_{1}t_{1}t_{1}}+\tfrac{3}{8}%
(q-t_{2})q_{t_{1}}q_{t_{1}t_{1}t_{1}} \\
&&-\frac{1}{2}(q-\tfrac{9}{8}t_{2})(q_{t_{1}t_{1}})^{2}-\tfrac{3}{8}%
(q_{t_{1}})^{2}q_{t_{1}t_{1}}-[\tfrac{15}{4}q^{3}-12t_{2}q^{2}+(\tfrac{51}{4}%
t_{2}^{2}+\tfrac{5}{4}t_{1})q-\tfrac{3}{2}t_{1}t_{2}-\tfrac{9}{2}%
t_{2}^{3}]q_{t_{1}t_{1}} \\
&&-\tfrac{3}{4}t_{1}(q_{t_{1}})^{2}-\tfrac{9}{2}q^{5}+\tfrac{99}{4}%
t_{2}q^{4}-(54t_{2}^{2}+3t_{1})q^{3}+(\tfrac{21}{2}t_{1}t_{2}+\tfrac{117}{2}%
t_{2}^{2})q^{2}-(12t_{1}t_{2}^{2}+\tfrac{63}{2}t_{2}^{4}+\tfrac{1}{2}%
t_{1}^{2})q \\
&&+\tfrac{27}{4}t_{2}^{5}+\tfrac{3}{4}t_{1}^{2}t_{2}+\tfrac{9}{2}%
t_{1}t_{2}^{3}+\tfrac{1}{16}-c,
\end{eqnarray*}%
which can be considered as the second equation from the standard $P_{34}$
hierarchy.

The isomonodromic Lax representation (\ref{5.18.6}) is as follows%
\begin{equation*}
L=\left( 
\begin{array}{cc}
-p_{1}\lambda +q_{2}p_{2} & \lambda ^{2}+q_{1}\lambda +q_{2} \\ 
\lambda ^{3}+(-q_{1}+3t_{2})\lambda
^{2}+(q_{1}^{2}-3q_{1}t_{2}-q_{2}+3t_{2}^{2}+t_{1})\lambda
+-q_{2}p_{2}^{2}+cq_{2}^{-1} & p_{1}\lambda -q_{2}p_{2}%
\end{array}%
\right) ,
\end{equation*}%
\begin{equation*}
U_{1}=\left( 
\begin{array}{cc}
0 & 1 \\ 
\lambda -2q_{1}+3t_{2} & 0%
\end{array}%
\right) ,\ \ \ \ U_{2}=\left( 
\begin{array}{cc}
-p_{1} & \lambda +q_{1} \\ 
\lambda ^{2}+(-q_{1}+3t_{2})\lambda
+q_{1}^{2}-2q_{2}-3t_{2}q_{1}+t_{1}+3t_{2}^{2} & p_{1}%
\end{array}%
\right) .
\end{equation*}

What is interesting, this system in flat coordinates $%
(x_{1},x_{2},p_{x_{1}},p_{x_{2}})$ on $R^{4}$, related with $(q,p)$
coordinates by a point transformation 
\begin{equation*}
q_{1}=-x_{1},\ \ \ q_{2}=-\tfrac{1}{4}x_{2}^{2},
\end{equation*}%
is the non-autonomous deformation of the famous Henon-Heiles system \cite%
{Fordy} consider in \cite{Honey}\ and in the complete version in \cite%
{Blaszak2019b}.

The third, three dimensional system from that hierarchy (\ref{5.18b}), i.e.
the deformed system from Example \ref{e2}, is generated by Hamiltonians%
\begin{align*}
H_{1}&
=h_{1}^{W}=h_{1}+5t_{3}V_{1}^{(5)}+(3t_{2}+10t_{3}^{2})V_{1}^{(4)}+(t_{1}+10t_{2}t_{3}+10t_{3}^{3})V_{1}^{(3)},
\\
H_{2}&
=h_{2}^{W}=h_{2}+p_{2}+5t_{3}V_{2}^{(5)}+(3t_{2}+10t_{3}^{2})V_{2}^{(4)}+(t_{1}+10t_{2}t_{3}+10t_{3}^{3})V_{2}^{(3)},
\\
H_{3}&
=h_{3}^{W}+t_{2}h_{1}^{W}=h_{3}+2p_{1}+q_{1}p_{2}+5t_{3}V_{3}^{(5)}+(3t_{2}+10t_{3}^{2})V_{3}^{(4)}+(t_{1}+10t_{2}t_{3}+10t_{3}^{3})V_{3}^{(3)}+t_{2}h_{1}^{W},\ \ \ \ \ \ \ \ \ \ \ \ 
\end{align*}%
where functions $h_{1},h_{2}$ and $h_{3}$ are given in Example \ref{e2}. The
related Hamiltonian vector fields $Y_{r}=\pi dH_{r}$, $r=1,2,3$ fulfill
Frobenius conditions (\ref{5.3}) and evolution equations $\xi _{t_{r}}=Y_{r}$
have the following isomonodromic Lax representations%
\begin{equation*}
L=\left( 
\begin{array}{cc}
-p_{2}\lambda ^{2}-(p_{1}+q_{1}p_{2})\lambda +q_{3}p_{3} & \lambda
^{3}+q_{1}\lambda ^{2}+q_{2}\lambda +q_{3} \\ 
L_{21} & -L_{11}%
\end{array}%
\right) ,
\end{equation*}%
\begin{eqnarray*}
L_{21} &=&\lambda ^{4}-(q_{1}-5t_{3})\lambda
^{3}-(V_{1}^{(4)}+5t_{3}q_{1}-3t_{2}-10t_{3}^{2})\lambda
^{2}-[V_{1}^{(5)}+p_{3}^{2}+5t_{3}V_{1}^{(4)}+(3t_{2}+10t_{3}^{2})q_{1} \\
&&-(t_{1}+10t_{2}t_{3}+10t_{3}^{3})]\lambda -q_{3}p_{3}^{2}+cq_{3}^{-1},
\end{eqnarray*}%
\begin{equation*}
\end{equation*}%
\begin{equation*}
U_{1}=\left( 
\begin{array}{cc}
0 & 1 \\ 
\lambda -2q_{1}+5t_{3} & 0%
\end{array}%
\right) ,\ \ \ U_{2}=\left( 
\begin{array}{cc}
-p_{2} & \lambda +q_{1} \\ 
\lambda ^{2}-(q_{1}-5t_{3})\lambda
+V_{1}^{(4)}-q_{2}-5t_{3}q_{1}+3t_{2}+10t_{3}^{2} & p_{2}%
\end{array}%
\right) ,
\end{equation*}%
\begin{equation*}
\end{equation*}%
\begin{equation*}
U_{3}=\left( 
\begin{array}{cc}
\begin{array}{c}
-p_{2}\lambda -q_{1}p_{2}-p_{1} \\ 
\left. {}\right.%
\end{array}
& 
\begin{array}{c}
\lambda ^{2}+q_{1}\lambda +q_{2}+t_{2} \\ 
\left. {}\right.%
\end{array}
\\ 
\begin{array}{c}
\lambda ^{3}-(q_{1}-5t_{3})\lambda
^{2}-(V_{1}^{(4)}+5t_{3}q_{1}-4t_{2}-10t_{3}^{2})\lambda \\ 
-p_{2}^{2}-V_{1}^{(5)}-q_{3}-5t_{3}V_{1}^{(4)}-(5t_{2}+10t_{3}^{2})q_{1}+t_{1}+15t_{2}t_{3}+10t_{3}^{3}%
\end{array}
& p_{2}\lambda +q_{1}p_{2}+p_{1}%
\end{array}%
\right) .
\end{equation*}

The systematic construction of isomonodromic Lax representations for higher
dimensional members (\ref{5.18b}) is described by (\ref{5.18.4})-(\ref%
{5.18.8}).

Also in that case, some elements of similar $P_{34}$-hierarchy appeared in 
\cite{Gordoa},\ where stationary flows of equations like these from (\ref%
{6.7}), but with time independent coefficients, was derived.

\section{Non-homogenous KdV hierarchies and related non-autonomous
stationary systems \label{6}}

In Sections \ref{3} and \ref{4} we have demonstrated how to reconstruct the
KdV hierarchy and its stationary systems form the hierarchies of St\"{a}ckel
systems (\ref{2.9}) and (\ref{4.3}), respectively. Here, by the same method,
we will construct two different non-homogeneous KdV hierarchies and related
non-autonomous stationary systems directly from Painlev\'{e} deformations (%
\ref{5.18a}) and (\ref{5.18b}) of considered St\"{a}ckel systems.

We begin from $P_{I}$ hierarchy (\ref{5.18a}). For one-dimensional equation (%
\ref{5.19}) ($P_{I}$), after identification $t_{1}=x$ and substitution $q=%
\frac{1}{2}u$ we get $p=\frac{1}{4}u_{x}$ and 
\begin{equation}
0=\tfrac{1}{4}u_{xx}+\tfrac{3}{4}u^{2}+x+c,\ \ \ \   \label{6.1}
\end{equation}%
which is the integrated stationary flow of the following PDE%
\begin{equation}
u_{t_{2}}=\mathcal{K}_{2}+\sigma _{-1}=\partial _{x}(\gamma _{2}+x+c)\equiv
\pi _{0}(\gamma _{1,2}+x+c)=\mathcal{K}_{1,2}  \label{6.2}
\end{equation}%
from the KdV family. By the same substitution, the Painlev\'{e}-type
equations generated by Hamiltonians $H_{1}$ from family (\ref{5.18a}), for $%
n=2,3,4,...,$ are integrated stationary flows of the following hierarchy of
non-homogeneous KdV equations 
\begin{align}
u_{t_{3}}& =\mathcal{K}_{3}+\tfrac{3}{2}t_{2}\mathcal{K}_{1}+\sigma
_{-1}\equiv \mathcal{K}_{2,3}=\pi _{0}(\gamma _{2,3}+x+c)  \notag \\
&  \notag \\
u_{t_{4}}& =\mathcal{K}_{4}+\tfrac{5}{2}t_{3}\mathcal{K}_{2}+\tfrac{3}{2}%
t_{2}\mathcal{K}_{1}+\sigma _{-1}\equiv \mathcal{K}_{3,4}=\pi _{0}(\gamma
_{3,4}+x+c)  \notag \\
&  \notag \\
u_{t_{5}}& =\mathcal{K}_{5}+\tfrac{7}{2}t_{4}\mathcal{K}_{3}+\tfrac{5}{2}%
t_{3}\mathcal{K}_{2}+\tfrac{3}{2}(t_{2}+\tfrac{7}{4}t_{4}^{2})\mathcal{K}%
_{1}+\sigma _{-1}\equiv \mathcal{K}_{4,5}=\pi _{0}(\gamma _{4,5}+x+c)
\label{6.3} \\
&  \notag \\
u_{t_{6}}& =\mathcal{K}_{6}+\tfrac{9}{2}t_{5}\mathcal{K}_{4}+\tfrac{7}{2}%
t_{4}\mathcal{K}_{3}+\tfrac{5}{2}(t_{3}+\tfrac{9}{4}t_{5}^{2})\mathcal{K}%
_{2}+\tfrac{3}{2}(t_{2}+\tfrac{9}{2}t_{4}t_{5})\mathcal{K}_{1}+\sigma
_{-1}\equiv \mathcal{K}_{5,6}=\pi _{0}(\gamma _{5,6}+x+c)  \notag \\
& \vdots  \notag
\end{align}

Contrary to the autonomous case, for fixed $n$, the remaining Painlev\'{e}%
-type equations generated by Hamiltonians $H_{2},...,H_{n}$ from (\ref{5.18a}%
) do not reconstruct the lower order equations from the hierarchy (\ref{6.3}%
). Actually, the hierarchy of Painlev\'{e}-type systems generated by
Hamiltonians (\ref{5.18a}) is equivalent to the following hierarchy of KdV
non-autonomous stationary systems%
\begin{equation}
\begin{array}{ll}
\begin{array}{l}
n=1: \\ 
\smallskip%
\end{array}
& 
\begin{array}{l}
\ \ 0=\gamma _{2}+x+c\equiv \mathcal{\gamma }_{1,2}+x+c \\ 
\smallskip%
\end{array}
\\ 
\begin{array}{l}
n=2: \\ 
\smallskip%
\end{array}
& 
\begin{array}{l}
u_{t_{2}}=\mathcal{K}_{2}\equiv \mathcal{K}_{2,2}=\pi _{0}\gamma _{2,2} \\ 
\ \ \ 0=\gamma _{3}+\frac{3}{2}t_{2}\gamma _{1}+x+c\equiv \gamma _{2,3}+x+c
\\ 
\smallskip%
\end{array}
\\ 
\begin{array}{l}
n=3: \\ 
\smallskip%
\end{array}
& 
\begin{array}{l}
u_{t_{2}}=\mathcal{K}_{2}\equiv \mathcal{K}_{3,2}\vspace{0.1cm}=\pi
_{0}\gamma _{3,2} \\ 
u_{t_{3}}=\mathcal{K}_{3}+\frac{5}{2}t_{3}\mathcal{K}_{1}\equiv \mathcal{K}%
_{3,3}\vspace{0.1cm}=\pi _{0}\gamma _{3,3} \\ 
\ \ \ 0=\mathcal{\gamma }_{4}+\frac{5}{2}t_{3}\gamma _{2}+\tfrac{3}{2}t_{2}%
\mathcal{\gamma }_{1}+\frac{5}{4}t_{3}^{2}+x+c\equiv \mathcal{\gamma }%
_{3,4}+x+c \\ 
\smallskip%
\end{array}
\\ 
\begin{array}{l}
n=4: \\ 
\smallskip%
\end{array}
& 
\begin{array}{l}
u_{t_{2}}=\mathcal{K}_{2}\equiv \mathcal{K}_{4,2}\vspace{0.1cm}=\pi
_{0}\gamma _{4,2} \\ 
u_{t_{3}}=\mathcal{K}_{3}+\frac{7}{2}t_{4}\mathcal{K}_{1}\equiv \mathcal{K}%
_{4,3}\vspace{0.1cm}=\pi _{0}\gamma _{4,3} \\ 
u_{t_{4}}=\mathcal{K}_{4}+\frac{7}{2}t_{4}\mathcal{K}_{2}+\frac{7}{2}t_{3}%
\mathcal{K}_{1}\equiv \mathcal{K}_{4,4}\vspace{0.1cm}=\pi _{0}\gamma _{4,4}
\\ 
\ \ \ 0=\mathcal{\gamma }_{5}+\tfrac{7}{2}t_{4}\mathcal{\gamma }_{3}+\tfrac{5%
}{2}t_{3}\mathcal{\gamma }_{2}+\tfrac{3}{2}(t_{2}+\tfrac{7}{4}t_{4}^{2})%
\mathcal{\gamma }_{1}+\frac{7}{2}t_{3}t_{4}+x+c\equiv \mathcal{\gamma }%
_{4,5}+x+c \\ 
\smallskip%
\end{array}%
\end{array}
\label{6.4}
\end{equation}
\begin{equation*}
\begin{array}{ll}
\begin{array}{l}
n=5: \\ 
\smallskip%
\end{array}
& 
\begin{array}{l}
u_{t_{2}}=\mathcal{K}_{2}\equiv \mathcal{K}_{5,2}\vspace{0.1cm}=\pi
_{0}\gamma _{5,2} \\ 
u_{t_{3}}=\mathcal{K}_{3}+\frac{9}{2}t_{5}\mathcal{K}_{1}\equiv \mathcal{K}%
_{5,3}\vspace{0.1cm}=\pi _{0}\gamma _{5,3} \\ 
u_{t_{4}}=\mathcal{K}_{4}+\frac{9}{2}t_{5}\mathcal{K}_{2}+\frac{7}{2}t_{4}%
\mathcal{K}_{1}\equiv \mathcal{K}_{5,4}\vspace{0.1cm}=\pi _{0}\gamma _{5,4}
\\ 
u_{t_{5}}=\mathcal{K}_{5}+\frac{9}{2}t_{5}\mathcal{K}_{3}+\frac{9}{2}t_{4}%
\mathcal{K}_{2}+\frac{9}{2}(t_{3}+\frac{5}{4}t_{5}^{2})\mathcal{K}_{1}\equiv 
\mathcal{K}_{5,5}\vspace{0.1cm}=\pi _{0}\gamma _{5,5} \\ 
\ \ \ 0=\mathcal{\gamma }_{6}+\tfrac{9}{2}t_{5}\mathcal{\gamma }_{4}+\tfrac{7%
}{2}t_{4}\mathcal{\gamma }_{3}+\tfrac{5}{2}(t_{3}+\tfrac{9}{4}t_{5}^{2})%
\mathcal{\gamma }_{2}+\tfrac{3}{2}(t_{2}+\tfrac{9}{2}t_{4}t_{5})\mathcal{%
\gamma }_{1} \vspace{0.1cm} \\ 
\ \ \ \ \ \ \ +\frac{15}{8}t_{5}^{3}+\frac{9}{2}t_{3}t_{5}+\frac{7}{4}%
t_{4}^{2}+x+c\equiv \mathcal{\gamma }_{5,6}+x+c \\ 
\smallskip%
\end{array}
\\ 
\vdots & 
\end{array}%
\end{equation*}

\begin{lemma}
For arbitrary $n\in \mathbb{N}$, non-autonomous vector fields $\mathcal{K}%
_{n,r}$ fulfill Frobenius conditions 
\begin{equation}
\lbrack \mathcal{K}_{n,s},\mathcal{K}_{n,r}]+\frac{\partial \mathcal{K}_{n,r}%
}{\partial t_{s}}-\frac{\partial \mathcal{K}_{n,s}}{\partial t_{r}}=0,\ \ \
\ \ r,s=2,...,n+1.  \label{6.5}
\end{equation}
\end{lemma}

\begin{theorem}
The non-autonomous stationary system 
\begin{equation}
u_{t_{r}}=\mathcal{K}_{n,r}\ =\pi _{0}\gamma _{n,r}\ ,\ \ \ 0=\gamma
_{n,n+1}+x+c,\ \ \ \ \ \ r=2,...,n\ \ \   \label{6.5a}
\end{equation}%
has isomonodromic Lax representation 
\begin{equation}
\frac{d}{dt_{r}}U_{n,n+1}=[U_{n,r},U_{n,n+1}]+\frac{\partial U_{n,r}}{%
\partial \lambda },\ \ \ \ \ r=1,...,n,  \label{6.5b}
\end{equation}%
where $\frac{d}{dt_{r}}$\ is the evolutionary derivative (\ref{4.14a}) along
the $r$-th flow $\mathcal{K}_{n,r}$, 
\begin{equation}
U_{n,r}=V_{n,r}\ \ \ \ r=1,...,\kappa ,\ \ \ \ \ U_{n,r}=\sum_{j=1}^{r}\zeta
_{r,j}(t_{1},\dotsc ,t_{r-1})V_{n,j},\quad \zeta _{r,r}=1,\quad \text{for $%
r=\kappa +1,\dotsc ,n,$}  \label{6.5c}
\end{equation}%
\begin{equation}
V_{n,r}=\left( 
\begin{array}{cc}
-\frac{1}{2}\left( P_{n,r}\right) _{x} & P_{n,r} \\ 
P_{n,r}(\lambda -u)-\frac{1}{2}\left( P_{n,r}\right) _{xx} & \frac{1}{2}%
\left( P_{n,r}\right) _{x}%
\end{array}%
\right) ,\ \ \ P_{n,r}=\frac{1}{2}\sum_{i=0}^{r-1}\gamma _{n,i}\lambda
^{r-i-1},\ \ \ \gamma _{n,0}=\gamma _{0}  \label{6.5d}
\end{equation}%
and 
\begin{equation*}
U_{n,n+1}=V_{n,n+1}\text{ \ under constraint \ }\ 0=\gamma _{n,n+1}+x+c.
\end{equation*}
\end{theorem}

The proof follows from the isomonodromic Lax representation (\ref{5.18.2}), (%
\ref{5.18.4}) of Painlev\'{e} representation of (\ref{6.5a}) and relations (%
\ref{2.15a}) and (\ref{2.15aa}) for their autonomous counterparts.

The hierarchy of non-homogeneous KdV equations (\ref{6.3}) has the following
non-isospectral zero curvature representation 
\begin{equation}
\frac{d}{dt_{n}}V_{1}+\frac{\partial }{\partial \lambda }V_{1}-\frac{d}{dx}%
V_{n,n+1}+[V_{1},V_{n,n+1}]=0,  \label{6.5e}
\end{equation}%
as $V_{n,1}=V_{1}$.

Now, let us pass to the second non-autonomous KdV hierarchy of stationary
systems, constructed from the Painlev\'{e}-type systems (\ref{5.18b}). Again
for $n=1,$ differentiation of (\ref{5.22}) by $t,$ division by $2q$ and
substitution $t=x,\ q=\frac{1}{2}u+\frac{1}{2}x$ we get the stationary flow
of the following PDE%
\begin{equation}
u_{t_{2}}=\left( \tfrac{1}{4}\partial _{x}^{3}+\tfrac{1}{2}u\partial _{x}+%
\tfrac{1}{2}\partial _{x}\right) (\gamma _{1}+x)=\mathcal{K}_{2}+\sigma
_{0}\equiv \pi _{1}\gamma _{1,1}=\mathcal{K}_{1,2}  \label{6.6}
\end{equation}%
from the KdV family. For $n=2,3,4,...,$ by the substitution $t_{1}=x,\ q_{1}=%
\frac{1}{2}u+\frac{2n-1}{2}t_{n}$, the Painlev\'{e}-type equations,
generated by Hamiltonians $H_{1}$ from (\ref{5.18b}) hierarchy, are
integrated stationary flows (with respect to $\pi _{1}$) of the following
hierarchy of non-homogeneous KdV equations 
\begin{align}
u_{t_{3}}& =\mathcal{K}_{3}+\tfrac{3}{2}t_{2}\mathcal{K}_{2}+\tfrac{3}{8}%
t_{2}^{2}\mathcal{K}_{1}+\sigma _{0}\equiv \mathcal{K}_{2,3}=\pi _{1}(%
\mathcal{\gamma }_{2}+\tfrac{3}{2}t_{2}\mathcal{\gamma }_{1}+\tfrac{3}{8}%
t_{2}^{2}\mathcal{\gamma }_{0}+x)=\pi _{1}\gamma _{2,2}  \notag \\
&  \notag \\
u_{t_{4}}& =\mathcal{K}_{4}+\tfrac{5}{2}t_{3}\mathcal{K}_{3}+\tfrac{3}{2}%
(t_{2}+\tfrac{5}{4}t_{3}^{2})\mathcal{K}_{2}+\tfrac{1}{2}(\tfrac{5}{2}%
t_{2}t_{3}+\tfrac{5}{8}t_{3}^{2})\mathcal{K}_{1}+\sigma _{0}\equiv \mathcal{K%
}_{3,4}  \notag \\
& =\pi _{1}\left[ \mathcal{\gamma }_{3}+\tfrac{5}{2}t_{3}\mathcal{\gamma }%
_{2}+\tfrac{3}{2}(t_{2}+\tfrac{5}{4}t_{3}^{2})\mathcal{\gamma }_{1}+\tfrac{1%
}{2}(\tfrac{5}{2}t_{2}t_{3}+\tfrac{5}{8}t_{3}^{2})\mathcal{\gamma }_{0}+x%
\right] =\pi _{1}\gamma _{3,3}  \notag \\
&  \notag \\
u_{t_{5}}& =\mathcal{K}_{5}+\tfrac{7}{2}t_{4}\mathcal{K}_{4}+\tfrac{5}{2}%
(t_{3}+\tfrac{7}{4}t_{4}^{2})\mathcal{K}_{3}+(\tfrac{3}{2}t_{2}+\tfrac{21}{4}%
t_{3}t_{4}+\tfrac{35}{16}t_{4}^{3})\mathcal{K}_{2}+(\tfrac{5}{8}t_{3}^{2}+%
\tfrac{7}{4}t_{2}t_{4}+\tfrac{35}{128}t_{4}^{4})\mathcal{K}_{1}+\sigma
_{0}\equiv \mathcal{K}_{4,5}  \notag \\
& =\pi _{1}\left[ \mathcal{\gamma }_{4}+\tfrac{7}{2}t_{4}\mathcal{\gamma }%
_{3}+\tfrac{5}{2}(t_{3}+\tfrac{7}{4}t_{4}^{2})\mathcal{\gamma }_{2}+(\tfrac{3%
}{2}t_{2}+\tfrac{21}{4}t_{3}t_{4}+\tfrac{35}{16}t_{4}^{3})\mathcal{\gamma }%
_{1}+(\tfrac{5}{8}t_{3}^{2}+\tfrac{7}{4}t_{2}t_{4}+\tfrac{35}{128}t_{4}^{4})%
\mathcal{\gamma }_{0}+x\right] =\pi _{1}\gamma _{4,4}  \notag \\
&  \label{6.7} \\
u_{t_{6}}& =\mathcal{K}_{6}+\tfrac{9}{2}t_{5}\mathcal{K}_{5}+(\tfrac{7}{2}%
t_{4}+\tfrac{63}{8}t_{5}^{2})\mathcal{K}_{4}+(\tfrac{5}{2}t_{3}+\tfrac{45}{4}%
t_{4}t_{5}+\tfrac{105}{16}t_{5}^{3})\mathcal{K}_{3}+(\tfrac{3}{2}t_{2}+%
\tfrac{21}{8}t_{4}^{2}+\tfrac{27}{4}t_{3}t_{5}  \notag \\
& \ \ \ \ +\tfrac{189}{16}t_{4}t_{5}^{2}+\tfrac{315}{128}t_{5}^{4})\mathcal{K%
}_{2}+(\tfrac{7}{4}t_{3}t_{4}+\tfrac{9}{4}t_{2}t_{5}+\tfrac{45}{16}%
t_{4}^{2}t_{5}+\tfrac{63}{16}t_{3}t_{5}^{2}+\tfrac{105}{32}t_{4}t_{5}^{3}+%
\tfrac{62}{256}t_{5}^{5})\mathcal{K}_{1}+\sigma _{0}\equiv \mathcal{K}_{5,6}
\notag \\
& =\pi _{1}\left[ \mathcal{\gamma }_{5}+\tfrac{9}{2}t_{5}\mathcal{\gamma }%
_{4}+(\tfrac{7}{2}t_{4}+\tfrac{63}{8}t_{5}^{2})\mathcal{\gamma }_{3}+(\tfrac{%
5}{2}t_{3}+\tfrac{45}{4}t_{4}t_{5}+\tfrac{105}{16}t_{5}^{3})\mathcal{\gamma }%
_{2}+(\tfrac{3}{2}t_{2}+\tfrac{21}{8}t_{4}^{2}+\tfrac{27}{4}t_{3}t_{5}\right.
\notag \\
& \ \ \ \ \ \ \ \ \left. +\tfrac{189}{16}t_{4}t_{5}^{2}+\tfrac{315}{128}%
t_{5}^{4})\mathcal{\gamma }_{1}+(\tfrac{7}{4}t_{3}t_{4}+\tfrac{9}{4}%
t_{2}t_{5}+\tfrac{45}{16}t_{4}^{2}t_{5}+\tfrac{63}{16}t_{3}t_{5}^{2}+\tfrac{%
105}{32}t_{4}t_{5}^{3}+\tfrac{63}{256}t_{5}^{5})\mathcal{\gamma }_{0}+x%
\right] =\pi _{1}\gamma _{5,5}  \notag \\
& \vdots  \notag
\end{align}

Again, contrary to the autonomous case, for fixed $n$, the remaining Painlev%
\'{e}-type equations generated by Hamiltonians $H_{2},...,H_{n}$ from (\ref%
{5.18b}) do not reconstruct the lower order equations from the hierarchy (%
\ref{6.7}). Actually, the hierarchy of Painlev\'{e}-type systems generated
by Hamiltonians (\ref{5.18b}) is equivalent to the following hierarchy of
non-autonomous KdV stationary systems%
\begin{equation}
\begin{array}{ll}
\begin{array}{l}
n=1: \\ 
\smallskip%
\end{array}
& 
\begin{array}{l}
\ \ \ 0=\tfrac{1}{2}\gamma _{1,1}(\gamma _{1,1})_{xx}-\tfrac{1}{4}[(\gamma
_{1,1})_{x}]^{2}+u\gamma _{1,1}^{2}+c \\ 
\smallskip%
\end{array}
\\ 
\begin{array}{l}
n=2: \\ 
\smallskip%
\end{array}
& 
\begin{array}{l}
u_{t_{2}}=\mathcal{K}_{2}+\frac{3}{2}t_{2}\mathcal{K}_{1}\equiv \mathcal{K}%
_{2,2}\vspace{0.1cm}=\pi _{1}\gamma _{2,1} \\ 
\ \ \ 0=\tfrac{1}{2}\gamma _{2,2}(\gamma _{2,2})_{xx}-\tfrac{1}{4}[(\gamma
_{2,2})_{x}]^{2}+u\gamma _{2,2}^{2}+c \\ 
\smallskip%
\end{array}
\\ 
\begin{array}{l}
n=3: \\ 
\smallskip%
\end{array}
& 
\begin{array}{l}
u_{t_{2}}=\mathcal{K}_{2}+\frac{5}{2}t_{3}\mathcal{K}_{1}\equiv \mathcal{K}%
_{3,2}\vspace{0.1cm}=\pi _{1}\gamma _{3,1} \\ 
u_{t_{3}}=\mathcal{K}_{3}+\frac{5}{2}t_{3}\mathcal{K}_{2}+(\frac{5}{2}t_{2}+%
\frac{15}{8}t_{3}^{2})\mathcal{K}_{1}\equiv \mathcal{K}_{3,3}=\pi _{1}\gamma
_{3,2}\ \vspace{0.1cm} \\ 
\ \ \ 0=\tfrac{1}{2}\gamma _{3,3}(\gamma _{3,3})_{xx}-\tfrac{1}{4}[(\gamma
_{3,3})_{x}]^{2}+u\gamma _{3,3}^{2}+c \\ 
\smallskip%
\end{array}
\\ 
\begin{array}{l}
n=4: \\ 
\smallskip%
\end{array}
& 
\begin{array}{l}
u_{t_{2}}=\mathcal{K}_{2}+\frac{7}{2}t_{4}\mathcal{K}_{1}\equiv \mathcal{K}%
_{4,2}=\pi _{1}\gamma _{4,1}\ \vspace{0.1cm} \\ 
u_{t_{3}}=\mathcal{K}_{3}+\frac{7}{2}t_{4}\mathcal{K}_{2}+(\frac{5}{2}t_{3}+%
\frac{35}{8}t_{4}^{2})\mathcal{K}_{1}\equiv \mathcal{K}_{4,3}\vspace{0.1cm}%
=\pi _{1}\gamma _{4,2} \\ 
u_{t_{4}}=\mathcal{K}_{4}+\frac{7}{2}t_{4}\mathcal{K}_{3}+(\frac{7}{2}t_{3}+%
\frac{35}{8}t_{4}^{2})\mathcal{K}_{2}+(\frac{7}{2}t_{2}+\frac{35}{4}%
t_{3}t_{4}+\frac{35}{16}t_{4}^{3})\mathcal{K}_{1}\equiv \mathcal{K}_{4,4}%
\vspace{0.1cm}=\pi _{1}\gamma _{4,3} \\ 
\ \ \ 0=\tfrac{1}{2}\gamma _{4,4}(\gamma _{4,4})_{xx}-\tfrac{1}{4}[(\gamma
_{4,4})_{x}]^{2}+u\gamma _{4,4}^{2}+c \\ 
\smallskip%
\end{array}
\\ 
n=5: & 
\begin{array}{l}
u_{t_{2}}=\mathcal{K}_{2}+\frac{9}{2}t_{5}\mathcal{K}_{1}\equiv \mathcal{K}%
_{5,2}\vspace{0.1cm}=\pi _{1}\gamma _{5,1} \\ 
u_{t_{3}}=\mathcal{K}_{3}+\frac{9}{2}t_{5}\mathcal{K}_{2}+(\frac{7}{2}t_{4}+%
\frac{63}{8}t_{5}^{2})\mathcal{K}_{1}\equiv \mathcal{K}_{5,3}\vspace{0.1cm}%
=\pi _{1}\gamma _{5,2} \\ 
u_{t_{4}}=\mathcal{K}_{4}+\frac{9}{2}t_{5}\mathcal{K}_{3}+(\frac{7}{2}t_{4}+%
\frac{63}{8}t_{5}^{2})\mathcal{K}_{2}+(\frac{7}{2}t_{3}+\frac{45}{4}%
t_{4}t_{5}+\frac{105}{16}t_{5}^{3})\mathcal{K}_{1}\equiv \mathcal{K}_{5,4}%
\vspace{0.1cm}=\pi _{1}\gamma _{5,3} \\ 
u_{t_{5}}=\mathcal{K}_{5}+\frac{9}{2}t_{5}\mathcal{K}_{4}+(\frac{9}{2}t_{4}+%
\frac{63}{8}t_{5}^{2})\mathcal{K}_{3}+(\frac{9}{2}t_{3}+\frac{63}{4}%
t_{4}t_{5}+\frac{105}{16}t_{5}^{3})\mathcal{K}_{2}\vspace{0.1cm} \\ 
\ \ \ \ \ \ \ \ +(\frac{9}{2}t_{2}+\frac{45}{8}t_{4}^{2}+\frac{63}{4}%
t_{3}t_{5}+\frac{315}{16}t_{4}t_{5}^{2}+\frac{315}{128}t_{5}^{4})\mathcal{K}%
_{1}\equiv \mathcal{K}_{5,5}\vspace{0.1cm}=\pi _{1}\gamma _{5,4} \\ 
\ \ \ 0=\tfrac{1}{2}\gamma _{5,5}(\gamma _{5,5})_{xx}-\tfrac{1}{4}[(\gamma
_{5,5})_{x}]^{2}+u\gamma _{5,5}^{2}+c%
\end{array}
\\ 
\vdots & 
\end{array}
\label{6.8}
\end{equation}

As in the previous case, for arbitrary $n\in \mathbb{N}$, non-autonomous
vector fields $\mathcal{K}_{n,r}\ ,\ r=2,...,n+1$ fulfill Frobenius
conditions (\ref{6.5}).

The non-autonomous stationary system 
\begin{equation*}
u_{t_{r}}=\mathcal{K}_{n,r}\ =\pi _{1}\gamma _{n,r-1}\ ,\ \ \ \ 0=\tfrac{1}{2%
}\gamma _{n,n}(\gamma _{n,n})_{xx}-\tfrac{1}{4}[(\gamma
_{n,n})_{x}]^{2}+u\gamma _{n,n}^{2}+c,\ \ \ \ \ r=2,...,n
\end{equation*}%
has isomonodromic Lax representation (\ref{6.5b})-(\ref{6.5d}), where now 
\begin{equation*}
U_{n,n+1}=V_{n,n+1}\text{ \ under constraint \ }\ 0=\tfrac{1}{2}\gamma
_{n,n}(\gamma _{n,n})_{xx}-\tfrac{1}{4}[(\gamma _{n,n})_{x}]^{2}+u\gamma
_{n,n}^{2}+c.
\end{equation*}%
Besides, the hierarchy of non-homogeneous KdV equations (\ref{6.7}) has
again the non-isospectral zero curvature representation in the form (\ref%
{6.5e}).

\section{Conclusions}

For the KdV hierarchy (\ref{2.2}) the related stationary systems (\ref{s1})
have two different representations (\ref{s3}) and (\ref{s4}), being
particular St\"{a}ckel systems. On the other hand, starting from the family
of such St\"{a}ckel systems, one can reconstruct related stationary systems (%
\ref{s1}) and then the whole KdV hierarchy (\ref{2.2}). In this article we
have performed the same procedure for Painlev\'{e} deformations of
considered St\"{a}ckel systems. In consequence we have constructed two
non-autonomous families of KdV hierarchies 
\begin{equation*}
u_{t_{n,r}}=\mathcal{K}_{n,r}(t)=\pi _{0}\gamma _{n,r}(t),\ \ \ \ \
u_{t_{n,n+1}}=\mathcal{K}_{n,n+1}(t)=\pi _{0}(\gamma _{n,n+1}(t)+x),\ \ \ \
r=2,...,n,\ \ \ \ n\in \mathbb{N}
\end{equation*}%
and%
\begin{equation*}
u_{\tau _{n,r}}=\mathcal{K}_{n,r}(\tau )=\pi _{1}\gamma _{n,r-1}(\tau ),\ \
\ \ \ u_{\tau _{n,n+1}}=\mathcal{K}_{n,n+1}(\tau )=\pi _{1}(\gamma
_{n,n}(\tau )+x),\ \ \ \ r=2,...,n,\ \ \ \ n\in \mathbb{N}
\end{equation*}%
with\ related non-autonomous stationary systems 
\begin{equation*}
u_{t_{n,r}}=\mathcal{K}_{n,r}(t)=\pi _{0}\gamma _{n,r}(t),\ \ \ \ \ 0=\gamma
_{n,n+1}(t)+x+c,\ \ \ \ r=2,...,n,\ \ \ \ n\in \mathbb{N}
\end{equation*}%
and 
\begin{equation*}
u_{\tau _{n,r}}=\mathcal{K}_{n,r}(\tau )=\pi _{1}\gamma _{n,r-1}(\tau ),\ \
\ \ \ 0=\tfrac{1}{2}\overline{\gamma }_{n,n}(\overline{\gamma }_{n,n})_{xx}-%
\tfrac{1}{4}[(\overline{\gamma }_{n,n})_{x}]^{2}+u\overline{\gamma }%
_{n,n}^{2}+c,\ \ \ r=2,...,n,\ \ \ \ n\in \mathbb{N}
\end{equation*}%
where $\ \overline{\gamma }_{n,n}=\gamma _{n,n}(\tau )+x,\ $having
respective Painlev\'{e} representations, considered in Section \ref{5}.

\end{document}